\documentclass[a4paper,english,aps, preprint]{revtex4}%
\usepackage[T1]{fontenc}
\usepackage[latin9]{inputenc}
\usepackage{amsmath}
\usepackage{graphicx}
\usepackage{esint}
\usepackage{babel}
\usepackage{amsfonts}
\usepackage{amssymb}%

\makeatother

\begin{document}
\today
\preprint{TPJU-3/2009}
\title{Photon Distribution Amplitudes in nonlocal chiral quark model}
\author{Piotr Kotko}
\email{kotko@th.if.uj.edu.pl}
\author{Michal Praszalowicz}
\email{michal@if.uj.edu.pl}
\affiliation{M. Smoluchowski Institute of Physics, Jagiellonian University, ul. Reymonta 4,
30-059 Krak{\'{o}}w, Poland}

\begin{abstract}
Photon distribution amplitudes up to twist four are calculated within the
nonlocal chiral quark model with a simple pole ansatz for momentum dependence
of the constituent quark mass. Calculations are performed using modified
electromagnetic vector current in order to satisfy Ward identities. Quark
condensate and magnetic susceptibility of the QCD vacuum entering definitions
of the distribution amplitudes are computed and compared with existing
phenomenological estimates. Both real and off-shell photons are considered and
relevant form factors are calculated. Our results are analytical up to the
numerical solution of certain algebraic equation.

\end{abstract}
\maketitle


\section{Introduction}

\label{intro}

In the present paper we calculate photon distribution amplitudes (DA) within a
low energy nonlocal model based on the instanton model of the QCD vacuum.
There are seven different photon DAs corresponding to the Dirac structure
probing the photon and to the light-cone twist (here we follow closely
definitions of ref. \cite{Ball:2002ps}). However in reality only one of them,
twist 2 tensor DA, can be accessed experimentally in hard exclusive processes
\cite{Braun:2002en,Rohrwild:2007yt,Pire,Szymanowski:2009tc} (for experimental
overview see ref. \cite{Ashery:2006zw}). Higher twist amplitudes are
suppressed in hard processes and vector twist 2 DA decouples for real photons.
However, the interest in the remaining photon DAs is not purely academical.
They are normalized through low energy observables such as quark condensate --
$\left\langle \bar{\psi}\psi\right\rangle $, magnetic susceptibility --
$\chi_{m}$ and mixed quark-gluon condensate -- $f_{3\gamma}$ that are of
importance for our understanding of the properties of the QCD vacuum. Only the
calculation of all photon DAs provides a test of the whole approach and may
prove its consistency.

In the present work we employ a nonlocal generalization of the semibosonized
Nambu Jona-Lasinio (NJL) model with momentum dependent constituent quark mass
$M(p)$ (which will be denoted as $M_{p}$) which follows from the instanton
model of the QCD vacuum \cite{Diakonov:1983hh,Shuryak:1984kp}. This model in
the present version has been previously used to calculate pion
\cite{Praszalowicz:2001wy,Praszalowicz:2001pi,Bzdak:2003qe} and kaon
\cite{Nam:2006au,Nam:2006mb,Nam:2006sx} distribution amplitudes, two pion DAs
\cite{Praszalowicz:2003pr}, pion-to-photon transition distribution amplitudes
\cite{Kotko:2008gy} and also twist 2 tensor photon DA \cite{Petrov:1998kg}
(see also \cite{Praszalowicz:2001wy}). However a complete analysis of all
seven photon DAs has been, to the best of our knowledge, conducted only in
ref. \cite{Dorokhov:2006qm} in a model similar to ours with, however, results
that in some respects are different than the ones obtained in the present paper.

One of the obvious problems that arises when one considers momentum dependent
fermion mass is the nonconservation of the naive vector current containing
only $\gamma^{\mu}$ Dirac matrix. There are many proposals in the literature
how to extend vector current to satisfy electromagnetic Ward
identities. None of them is unique, since current conservation does not fix
the transverse part of the modified vertex. One of the simplest extensions of
this type discussed already many years ago in
refs. \cite{Ball:1980ay,Frank:1994gc} and employed also in
ref. \cite{Dorokhov:2006qm}, consists in the following substitution%
\begin{equation}
\gamma^{\mu}\rightarrow\gamma^{\mu}-\frac{k^{\mu}+(k-p)^{\mu}}{k^{2}%
-(k-p)^{2}}\left(  M_{k}-M_{k-p}\right)  . \label{eq:vertex}%
\end{equation}
Extension (\ref{eq:vertex}) follows from the assumption concerning
both Ward identities and analytical structure of the modified
vertex that is required to match perturbative expansion.
In principle one could add to (\ref{eq:vertex}) terms proportional
to $r^{\mu}$ where $r\cdot p=0$, and the Ward identities would be satisfied. In
refs. \cite{Bowler:1994ir,Plant:1997jr} and also in \cite{Broniowski:1999bz}
another modification has been advocated; here the ambiguity is connected to a
freedom of choosing the integration path that defines the nonlocal vertex. In
view of this ambiguous situation we have decided to use the simplest possible
extension of ref. \cite{Ball:1980ay} given by eq. (\ref{eq:vertex}).

Unlike the pion or the $\rho$ meson the photon has a dual nature being both
point-like and composite at the same time. Therefore in order to calculate
photon DAs that describe nonperturbative quark-antiquark structures, one has
to subtract the perturbative part. In order to avoid ambiguities this
procedure has to be well defined. In our case, since we work in the chiral
limit, we subtract the perturbative part only from these photon DAs that are
nonzero for massless quarks. These are:  vector and axial DAs which are also
UV divergent. Therefore the subtraction of the perturbative part renormalizes
these DAs ensuring their finiteness. At the same time it introduces a term
proportional to $\ln(-P^{2})$ that develops imaginary part for positive
virtualities. This is the reflection of the fact that the photon can decay
into free massless quarks in the chirally even channels. Throughout this paper
we plot DAs both for negative and positive photon virtualities, in the latter
case we take just the real part if the imaginary part exists. We also display
momentum dependence of the pertinent decay constants that are characterized by
dimensionless functions $F_{T,V,A}(P^{2})$ for brevity referred to as form factors.

Finally let us make a technical remark concerning loop integrals over $d^{4}k
$ with $k^{+}=uP^{+}$ component fixed by the $\delta$ function. Such
integrals, depending on the tensor structure may contain "singular" pieces
proportional to $\delta(u)$ and $\delta(u-1)$ or even the derivatives of the
$\delta$ functions. This is the case for higher twist photon DAs only. We
discuss this in more detail in sect.~\ref{loopi}, here we just want to point
out that higher twist DAs are in fact distributions rather than ordinary
functions. Quite importantly, the $\delta$ function contribution is always
accompanied by a regular piece that together with the $\delta$ piece
integrates to zero for any $P^{2}$. This allows to define a properly
normalized regular part and a singular part of DA of zero norm.

In the next section we introduce kinematical variables and define photon
distribution amplitudes. In sect.~\ref{nonlo} we describe the main features of
our model specifying the ansatz for the momentum dependence of the constituent
quark mass. We fix model parameters requiring that the experimental value of
the pion decay constant $F_{\pi}=93$~MeV is reproduced. To this end we use
model formula given in eq. (\ref{eq:Fpi}). Next, in sect.~\ref{loopi}, we
describe techniques used to calculate loop integrals with momentum dependent
constituent quark mass. We pay special attention to Lorentz invariance and
show how the end point singularities proportional to the Dirac $\delta$
functions arise. Main results are presented in sect.~\ref{photo}. First we
calculate dimensional constants entering definitions of the DAs
(\ref{eq:def_tensor})--(\ref{eq:def_axial}), namely quark condensate, magnetic
susceptibility and yet another constant, called $f_{3\gamma}$. We obtain
numerical results that agree with the "experimental" values known from
phenomenology. Finally in sections \ref{leadi} and \ref{highe} we present our
main results plotting different DAs and discussing their properties.

Our results can be briefly summarized as follows. Leading twist amplitudes are
rather insensitive to model parameters, whereas higher twist amplitudes
exhibit much stronger dependence, moreover some of them contain $\delta$
functions. We also show the influence of the nonlocal part of the photon
vertex (\ref{eq:vertex}) on the shape of photon DAs. For some DAs it is
rather unimportant, for the other ones it is absolutely crucial. More
discussion and comparison with other models is given in sect.~\ref{summa}.
Technical details are collected in appendices.

\section{Definitions and kinematics}

\label{defin}

Photon distribution amplitudes are defined through matrix elements of the
nonlocal quark-antiquark billinears between vacuum and one photon state. Quark
operators are assumed to be on the light cone separated by the distance
$2\lambda$. In the following we use the light-cone coordinates defined by two
light like vectors $n^{\mu}$ and $\tilde{n}^{\mu}$ such that $n^{\mu}=\left(
1,0,0,-1\right)  $ and $\tilde{n}^{\mu}=\left(  1,0,0,1\right)  $. In this
basis any four-vector $v^{\mu}$ can be decomposed into $v^{+}$ and $v^{-}$
components%
\begin{equation}
v^{\mu}=v^{+}\frac{\tilde{n}^{\mu}}{2}+v^{-}\frac{n^{\mu}}{2}+v_{\bot}^{\mu}.
\label{eq:lightcoord1}%
\end{equation}
Scalar product can be written as%
\begin{equation}
u\cdot v=\frac{1}{2}u^{+}v^{-}+\frac{1}{2}u^{-}v^{+}-\vec{u}_{\bot}\cdot
\vec{v}_{\bot}. \label{eq:lightcoord2}%
\end{equation}
We shall work in the system where the photon momentum is expressed as%
\begin{equation}
P^{\mu}=P^{+}\frac{\tilde{n}^{\mu}}{2}+\frac{P^{2}}{P^{+}}\frac{n^{\mu}}{2}.
\label{eq:kin1}%
\end{equation}
Decomposition of the polarization vector reads%
\begin{equation}
\varepsilon^{\mu}=\varepsilon^{+}\frac{\tilde{n}^{\mu}}{2}+\varepsilon
^{-}\frac{n^{\mu}}{2}+\varepsilon_{\bot}^{\mu},\quad\text{where}%
\quad\varepsilon_{\bot}^{\mu}\varepsilon_{\bot\,\mu}=-\vec{\varepsilon}_{\bot
}\cdot\vec{\varepsilon}_{\bot}=-1. \label{eq:kin2}%
\end{equation}
Since $P\cdot\varepsilon=0$ we have the relation%
\begin{equation}
\varepsilon^{-}=-\frac{P^{2}}{\left(  P^{+}\right)  ^{2}}\varepsilon^{+}.
\label{eq:kin3}%
\end{equation}
For real photon we obviously have $\varepsilon^{+}=\varepsilon\cdot n=0$ and
consequently $\varepsilon^{-}=0$ as well.

Depending on different tensor nature of the bilocal operators, we can define
tensor%
\begin{align}
\left\langle 0\left\vert \overline{\psi}\left(  \lambda n\right)
\sigma^{\alpha\beta}\psi\left(  -\lambda n\right)  \right\vert \gamma\left(
P,\varepsilon\right)  \right\rangle  &  =i\frac{e}{2}\left\langle \bar{\psi
}\psi\right\rangle F_{T}\left(  P^{2}\right) \nonumber\\
&  \Bigg\{\left(  \varepsilon_{\bot}^{\alpha}\tilde{n}^{\beta}-\varepsilon
_{\bot}^{\beta}\tilde{n}^{\alpha}\right)  P^{+}\chi_{m}\int_{0}^{1}%
du\,e^{i\xi\lambda P^{+}}\,\phi_{T}\left(  u,P^{2}\right) \nonumber\\
&  \frac{1}{P^{+}}\left(  \tilde{n}^{\alpha}n^{\beta}-\tilde{n}^{\beta
}n^{\alpha}\right)  \varepsilon^{+}\,\int_{0}^{1}du\,e^{i\xi\lambda P^{+}%
}\,\psi_{T}\left(  u,P^{2}\right) \nonumber\\
&  \frac{1}{P^{+}}\left(  \varepsilon_{\bot}^{\alpha}n^{\beta}-\varepsilon
_{\bot}^{\beta}n^{\alpha}\right)  \int_{0}^{1}du\,e^{i\xi\lambda P^{+}}%
\,h_{T}\left(  u,P^{2}\right)  \Bigg\} \label{eq:def_tensor}%
\end{align}
vector%
\begin{align}
\left\langle 0\left\vert \overline{\psi}\left(  \lambda n\right)  \gamma^{\mu
}\psi\left(  -\lambda n\right)  \right\vert \gamma\left(  P,\varepsilon
\right)  \right\rangle  &  =ef_{3\gamma}F_{V}\left(  P^{2}\right) \nonumber\\
&  \Bigg\{\frac{1}{2}\tilde{n}^{\mu}\varepsilon^{+}\int_{0}^{1}du\,e^{i\xi
\lambda P^{+}}\,\phi_{V}\left(  u,P^{2}\right) \nonumber\\
&  +\varepsilon_{\bot}^{\mu}\,\int_{0}^{1}du\,e^{i\xi\lambda P^{+}}\,\psi
_{V}\left(  u,P^{2}\right) \nonumber\\
&  -\frac{1}{2}\frac{P^{2}}{\left(  P^{+}\right)  ^{2}}n^{\mu}\varepsilon
^{+}\int_{0}^{1}du\,e^{i\xi\lambda P^{+}}\,h_{V}\left(  u,P^{2}\right)
\Bigg\} \label{eq:def_vector}%
\end{align}
and axial vector%
\begin{align}
\left\langle 0\left\vert \overline{\psi}\left(  \lambda n\right)  \gamma^{\mu
}\gamma_{5}\psi\left(  -\lambda n\right)  \right\vert \gamma\left(
P,\varepsilon\right)  \right\rangle  &  =\frac{1}{2}ef_{3\gamma}F_{A}\left(
P^{2}\right) \nonumber\\
&  \epsilon_{\mu\nu\alpha\beta}\varepsilon_{\bot}^{\nu}\tilde{n}^{\alpha
}n^{\beta}P^{+}\lambda\int_{0}^{1}du\,e^{i\xi\lambda P^{+}}\psi_{A}\left(
u,P^{2}\right)  \label{eq:def_axial}%
\end{align}
distribution amplitudes. For compactness we used notation $\xi=2u-1$ where $u$
is longitudinal fraction of the quark momentum and dropped Wilson lines
$\left[  -\lambda n,\lambda n\right]  $ that ensure gauge invariance of the
nonlocal operators. In the light-cone gauge $A\cdot n=0$ and hence $\left[
-\lambda n,\lambda n\right]  =1$. Moreover, since we use an effective model
where gluonic fields are integrated out, Wilson lines corresponding
to gluon fields never appear.

Our definitions follow closely those of refs. ~\cite{Ball:2002ps} and
\cite{Dorokhov:2006qm}, however we need only one $P^{2}-$dependent
dimensionless form factor for each tensor structure: $F_{T}\left(
P^{2}\right)  $, $F_{V}\left(  P^{2}\right)  $ and $F_{A}\left(  P^{2}\right)
$ where subscripts $T,\,V$ and $A$ stay for vector, tensor and axial vector,
respectively. Constant $\chi_{m}$ is the magnetic susceptibility of the quark
condensate $\left\langle \bar{\psi}\psi\right\rangle $, and $f_{3\gamma}$
corresponds to the axial mixed quark-gluon condensate. They provide natural
mass scales for distribution amplitudes. Analytical expressions for
$\left\langle \bar{\psi}\psi\right\rangle $, $\chi_{m}$ and $f_{3\gamma}$, and
for the form factors can be retrieved from the matrix elements of local
operators:%
\begin{align}
\left\langle 0\left\vert \overline{\psi}\left(  0\right)  \sigma^{\alpha\beta
}\psi\left(  0\right)  \right\vert \gamma\left(  P,\varepsilon\right)
\right\rangle  &  =ie\left\langle \bar{\psi}\psi\right\rangle \chi_{m}%
F_{T}\left(  P^{2}\right)  \left(  \varepsilon^{\alpha}P^{\beta}%
-\varepsilon^{\beta}P^{\alpha}\right)  ,\label{eq:tens_ff}\\
\left\langle 0\left\vert \overline{\psi}\left(  0\right)  \gamma^{\mu}%
\psi\left(  0\right)  \right\vert \gamma\left(  P,\varepsilon\right)
\right\rangle  &  =ef_{3\gamma}F_{V}\left(  P^{2}\right)  \varepsilon^{\mu
},\label{eq:vect_ff}\\
\left.  \frac{d}{d\lambda}\left\langle 0\left\vert \overline{\psi}\left(
-\lambda n\right)  \gamma^{\mu}\gamma_{5}\psi\left(  \lambda n\right)
\right\vert \gamma\left(  P,\varepsilon\right)  \right\rangle \right\vert
_{\lambda=0}  &  =ef_{3\gamma}F_{A}\left(  P^{2}\right)  \epsilon_{\mu
\nu\alpha\beta}\varepsilon^{\nu}P^{\alpha}n^{\beta}. \label{eq:ax_ff}%
\end{align}

Equations \eqref{eq:def_tensor}-\eqref{eq:def_axial} define photon
distribution amplitudes that can be classified according to the kinematical
light-cone twist. We have distributions of twist-2: $\phi_{T}$, $\phi_{V}$, of
twist-3: $\psi_{T}$, $\psi_{V}$, $\psi_{A}$ and of twist-4: $h_{T}$, $h_{V}$.
This can be easily seen by inspecting eqs. 
\eqref{eq:def_tensor}-\eqref{eq:def_axial}, since the  twist counting actually
reduces to counting the powers of $P^{+}$. Notice that in the case of axial
distribution the power of $P^{+}$ equals $1$, what would correspond to
twist-2, however additionally there is a path stretch $\lambda$ with inverse
mass dimensionality that makes $\psi_{A}$ to have twist 3 rather than 2.

Constants $\chi_{m}$, $\left\langle \bar{\psi}\psi\right\rangle $ and
$f_{3\gamma}$ are chosen in such a way that the following normalization
conditions are satisfied%
\begin{align}
\int_{0}^{1}\,\phi_{T}\left(  u,P^{2}\right)  du  &  =1,\quad\int_{0}%
^{1}\,\psi_{T}\left(  u,P^{2}\right)  du=\chi_{m}P^{2},\quad\int_{0}%
^{1}\,h_{T}\left(  u,P^{2}\right)  du=\chi_{m}P^{2},\label{eq:normaliz1}\\
\int_{0}^{1}\,\phi_{V}\left(  u,P^{2}\right)  du  &  =1,\quad\quad\int_{0}%
^{1}\,\psi_{V}\left(  u,P^{2}\right)  du=1,\quad\quad\quad\int_{0}^{1}%
\,h_{V}\left(  u,P^{2}\right)  du=1 \label{eq:normaliz2}%
\end{align}%
\begin{equation}
\quad\int_{0}^{1}\,\psi_{A}\left(  u,P^{2}\right)  du=1.\quad
\label{eq:normaliz3}%
\end{equation}

Note that due to the conservation of vector current $F_{V}(0)=0$. On
the other hand $F_{T}(0)=1$. Normalization of the axial form factor $F_{A}(0)$
is arbitrary and depends on the dimensional constant used in definition
(\ref{eq:def_axial}).

\section{Nonlocal chiral quark model}

\label{nonlo}

In order to calculate relevant matrix elements in the low energy domain we
shall use effective action based on the instanton vacuum theory
\cite{Diakonov:1983hh}. Its main feature is momentum dependent constituent
quark mass
\begin{equation}
M\left(  k\right)  =M\,F^{2}\left(  k\right)  \label{eq:massF}%
\end{equation}
appearing due to the chiral symmetry breaking. This dependence enters
not only into propagators, but serves as a nonlocal quark-meson coupling as
well. For zero momentum $M\left(  0\right)  \equiv M$ is of order of
$350\,\mathrm{MeV}$, while for $k\rightarrow\infty$ constituent quark mass vanishes
$M\left(  k\right)  \rightarrow0$.

One has to remember that the semibosonized NJL model, although devised to
describe chiral physics of Goldstone bosons, has been widely used to
incorporate baryons as chiral solitons both in local (for review see
\emph{e.g.} ref. \cite{Christov}) and non-local \cite{RipBroGo} cases.
Generally the results of these studies show that the soliton ceases to exist
for too small constituent quark mass $M$. The critical value of $M$ depends on
the details of the given model, however it is of the order of 300 MeV or a bit
less. Typical values of $M$ that fit well the hyperon spectrum may be as high
as 420 MeV \cite{Blotzetal}. In order to investigate dependence of photon DAs
on $M$ we use three distinct values of $M$: 300, 350 and 400 MeV.

Due to the momentum dependence of the quark mass, the naive vector current
$j^{\mu}=\bar{\psi}\gamma^{\mu}\psi$ violates electromagnetic Ward identities.
In order to fix this deficiency new nonlocal terms have to be added to
$j^{\mu}$. As already discussed in the Introduction there are several ways of
constructing extensions that make the current conserved. In the present paper
we use the simplest possible generalization of the vector current,
\cite{Ball:1980ay,Dorokhov:2006qm} replacing $\gamma^{\mu}$ by an effective
vertex of (\ref{eq:vertex}). One can easily check that electromagnetic Ward
identities are satisfied when (\ref{eq:vertex}) is used instead of
$\gamma^{\mu}$. Although eq. (\ref{eq:vertex}) introduces an extra pole inside
Feynman amplitudes, its residue, as we will explicitly show, is zero due to
the mass difference in the numerator. This generalization of the vector
current has been widely used in the literature also in the context of the
photon DAs \cite{Dorokhov:2006qm}.

Expression for the form factor $F_{\mathrm{inst}}\left(  k\right)  $ within
instanton vacuum model is known analytically in Euclidean space and is highly
nontrivial \cite{Diakonov:1983hh}. Therefore, in order to perform analytical
calculations directly in Minkowski space, we use the following formula
\cite{Praszalowicz:2001wy}%
\begin{equation}
F(k)=\left(  \frac{-\Lambda_{n}^{2}}{k^{2}-\Lambda_{n}^{2}+i\epsilon}\right)
^{n}, \label{Fkdef}%
\end{equation}
where $\Lambda_{n}$ is cutoff parameter adjusted for each $n$ in such a way,
that the experimental value of the pion decay constant is reproduced. 
For transparency we shall skip subscript $n$ and use $\Lambda$
rather than $\Lambda_n$ in the following.

Equation \eqref{Fkdef} reproduces reasonably well original shape $F_{\mathrm{inst}%
}\left(  k\right)  $ when continued to Euclidean momentum. It should be
however pointed out that expression \eqref{Fkdef} does not follow the
exponential asymptotics of $F_{\mathrm{inst}}\left(  k\right)  $
\cite{Diakonov:1983hh}. Parameter $n$ is introduced in
order to check sensitivity of our results to the shape of
$F\left(  k\right)$.

In order to fix the model parameter $\Lambda$ we use the following
Euclidean expression for the weak pion decay constant \cite{Bowler:1994ir}:%
\begin{equation}
F_{\pi}^{2}=\frac{N_{c}}{4\pi^{2}}\int_{0}^{\infty}d  k_{\mathrm{E}}%
^{2} \, k_{\mathrm{E}}^2  \frac{M^{2}\left(  k^2_{\mathrm{E}}\right)  -k_{\mathrm{E}}%
^{2}M\left(  k^2_{\mathrm{E}}\right)  M^{\prime}\left(  k^2_{\mathrm{E}}\right)
+k_{\mathrm{E}}^{4}M^{\prime}\left(  k^2_{\mathrm{E}}\right)  ^{2}}{\left(
k_{\mathrm{E}}^{2}+M^{2}\left(  k^2_{\mathrm{E}}\right)  \right)  ^{2}}
\label{eq:Fpi}%
\end{equation}
where $M^{\prime}(k^2_{\mathrm{E}})=
dM(k^2_{\mathrm{E}})/dk^2_{\mathrm{E}}$.
Notice that this formula differs from the Pagels-Stokar formula of
ref. ~\cite{Pagels:1979hd}. It has been also obtained in
ref. ~\cite{Bzdak:2003qe} from the PCAC relation in Minkowski space. Using
experimental value $F_{\pi}=93\,\mathrm{MeV}$ and (\ref{Fkdef}) we obtain
cutoff parameters listed in Table \ref{tab:0} for several choices of the
constituent quark mass $M$ and $n$. Analytical expression obtained within the
present model is given in appendix \ref{piond}. We remark at this point that
the cutoff parameter $\Lambda$ should not be confused with a typical
scale of the model, which for the instanton model is about
$600\,\mathrm{MeV}.$

\begin{table}[ptb]
\begin{centering}
\begin{tabular}{|c|c|}
\hline
\multicolumn{2}{|c|}{$M=300\,\mathrm{MeV}$}\tabularnewline
\hline
\hline
$n=1$ & $\Lambda=1016\,\mathrm{MeV}$\tabularnewline
\hline
$n=5$ & $\Lambda=2385\,\mathrm{MeV}$\tabularnewline
\hline
\hline
\multicolumn{2}{|c|}{$M=350\,\mathrm{MeV}$}\tabularnewline
\hline
\hline
$n=1$ & $\Lambda=836\,\mathrm{MeV}$\tabularnewline
\hline
$n=5$ & $\Lambda=1970\,\mathrm{MeV}$\tabularnewline
\hline
\hline
\multicolumn{2}{|c|}{$M=400\,\mathrm{MeV}$}\tabularnewline
\hline
\hline
$n=1$ & $\Lambda=721\,\mathrm{MeV}$\tabularnewline
\hline
$n=5$ & $\Lambda=1704\,\mathrm{MeV}$\tabularnewline
\hline
\end{tabular}
\par\end{centering}
\caption{Numerical values of the model parameters obtained using Birse-Bowler
formula for pion decay constant $F_{\pi}$.}%
\label{tab:0}%
\end{table}

\section{Loop integrals with momentum dependent mass}

\label{loopi}

In this section we present a brief sketch of our calculations underlying the
most important steps. Further technicalities are relegated to appendices. In
order to calculate the DA of interest -- denoted generically as $f(u)$ -- we
have to invert formulae (\ref{eq:def_tensor})--(\ref{eq:def_axial}) by
contracting them with appropriate 4-vectors and by performing Fourier
transform in $\lambda$. This results in the following formulae%
\begin{equation}
f(u)=\frac{-ie_{q}4P^{+}N_{c}}{C}\int\frac{d^{D}k}{\left(  2\pi\right)  ^{D}%
}\emph{T}_{\Gamma}(k,k-P)\,\delta(k^{+}-uP^{+})\label{loop}%
\end{equation}
where $C$ is the constant obtained by the contraction, $\Gamma$ is the
contracted tensor structure defining given amplitude and%
\begin{equation}
\emph{T}_{\Gamma}^{\,}(k,q)=\frac{1}{4}\operatorname{Tr}\left\{  \Gamma
\frac{1}{\not k-M_{k}+i\epsilon}\,\varepsilon_{\mu}\tilde{\gamma}^{\mu}\left(
k,k-P\right)  \frac{1}{(\not k-\not P)-M_{k-P}+i\epsilon}\right\}
\end{equation}
stands for the Dirac trace. Note that in the case of axial DA, because of
$\lambda$ standing in the l.h.s. of (\ref{eq:def_axial})%
\begin{equation}
f(u)=\frac{1}{2}
\left( \psi_{A}^{\prime}(u)+\psi_{A}(0)\delta(u)-\psi_{A}\delta(u-1)\right).
\end{equation}
Since some of the integrals can be UV divergent we shall work in
$D=4-2\epsilon$ dimensions.

Previous calculations using present nonlocal model were done by integration in
the light-cone coordinates, with special care concerning the integration
contour to ensure analyticity in $\Lambda_{n}$. Here we present another method
of performing such integrals based on the $\alpha$-representation for the
propagators. It is especially useful in the case of integrals appearing in
higher twist distributions, because of the end point delta-type singularities,
which are cumbersome to treat by integration in the light-cone coordinates.

The above complication can be well illustrated by considering a loop integral
of the type (\ref{loop}) for the numerator involving $k^{\mu}$. If not for the
$\delta$ function it would have been proportional to $P^{\mu},$ but because
$k^{+}-uP^{+}=n\cdot k-u\,n\cdot P$ we have%
\begin{equation}
\int\frac{d^{D}k}{\left(  2\pi\right)  ^{D}}\frac{k^{\mu}\,\mathcal{N}%
\,}{\left(  k^{2}-M_{k}^{2}+i\epsilon\right)  \left(  \left(  k-P\right)
^{2}-M_{k-P}^{2}+i\epsilon\right)  }\delta(k^{+}-uP^{+})\,=A(u)P^{\mu
}+B(u)n^{\mu}\label{loop1}%
\end{equation}
where $\mathcal{N}$ is some scalar function involving $n$, $\varepsilon$ and
$P$. There is an obvious condition following from Lorentz invariance%
\begin{equation}%
{\displaystyle\int\limits_{0}^{1}}
duB(u)=0.\label{LorentzB0}%
\end{equation}
However, as mentioned above and as shown explicitly in appendix \ref{light}
function $B(u)$ contains both regular piece and the piece with delta
functions: $\delta(u)$ and $\delta(u-1)$. Only the sum of both contributions
integrates over $du$ to zero. Note that this cancelation occurs for any
$P^{2}$. Since $n\cdot n=0$ and $\varepsilon_{\bot}\cdot n=0$, the delta
functions contribute only to the integrals of the $k^{-}$ component. The
integrals with tensor structure $k^{\mu}k^{\nu}$ are even more complicated,
since they involve derivatives of $\delta$ functions.

As it was already discussed in ref. ~\cite{Praszalowicz:2001wy} momentum mass
dependence given by \eqref{Fkdef} introduces a set of poles, whose positions
depend on parameter $\Lambda$. To this end it is convenient to introduce
dimensionless scaled variables%
\begin{equation}
\kappa=k/\Lambda_{n},\qquad p=P/\Lambda_{n},\qquad r=M/\Lambda_{n}
\label{eq:dimensionless_var}%
\end{equation}
and to define%
\begin{equation}
z_{1}=\left(  \kappa-p\right)  ^{2}-1+i\epsilon,\;z_{2}=\kappa^{2}%
-1+i\epsilon. \label{eq:z1z2_def}%
\end{equation}
Then the loop integral involving two propagators, like the one in.
eq. (\ref{loop1}), is transformed into%
\begin{equation}
\mathcal{I}=\Lambda^{D-5}\int\frac{d^{D}\kappa}{\left(  2\pi\right)  ^{D}%
}\,\delta\left(  \kappa\cdot n-up^{+}\right)  \frac{z_{1}^{4n}z_{2}%
^{4n}\mathcal{N}}{G\left(  z_{1}\right)  G\left(  z_{2}\right)  },
\label{loop2}%
\end{equation}
where the numerator $\mathcal{N}(z_{1},z_{2},\kappa\cdot n,\kappa\cdot
\tilde{n},\kappa\cdot\varepsilon_{\bot})$ depends on the DA considered. Here%
\begin{equation}
G\left(  z_{i}\right)  =z_{i}^{4n+1}+z_{i}^{4n}-r^{2}=\prod_{j=1}%
^{4n+1}\left(  z_{i}-\eta_{j}\right) \label{eq:Gz}
\end{equation}
corresponds to the propagator with momentum dependent mass (for $n=0$ it
reduces to the ordinary propagator in scaled variables) where the complex
numbers $\eta_{j}$ are roots of polynomial $G$ to be obtained numerically.

Next we decompose the inverse product of $G\left(  z_{1}\right)  G\left(
z_{2}\right)  $ into a sum of simple poles%
\begin{equation}
\frac{z_{1}^{M}z_{2}^{N}}{G\left(  z_{1}\right)  G\left(  z_{2}\right)  }%
=\sum_{i,j=1}^{4n+1}f_{i}f_{j}\frac{\eta_{i}^{M}\eta_{j}^{N}}{(z_{1}-\eta
_{i})(z_{2}-\eta_{j})}\;\text{for}\;M,N\leq4n\label{decomposition}%
\end{equation}
with%
\begin{equation}
f_{i}=\prod_{k=1,\,k\not =i}^{4n+1}\frac{1}{\eta_{i}-\eta_{k}}.\label{eq:f_i}%
\end{equation}
In this way integral (\ref{loop2}) is reduced to the sum of contributions
involving two propagators only. It is convenient to use the $\alpha
$-representation (exponential Schwinger representation) for the product of
propagators since also the $\delta$ function in (\ref{loop2}) can be written
as an exponent. Further calculations are rather standard and are summarized in
appendix \ref{light}. As a result we obtain analytical expressions given as
sums over roots $\eta_{i}$. Certain simplifications occur when we use the
following identity (which is true for any set of complex numbers $\{\eta
_{i}\}$ not only for the solutions of $G(\eta_{i})=0$):%
\begin{equation}
\sum_{i=1}^{4n+1}f_{i}\eta_{i}^{N}=%
\begin{cases}
0 & N<4n,\\
1 & N=4n.
\end{cases}
\label{eq:ident}%
\end{equation}
The proof of (\ref{eq:ident}) and other useful identities can be found in
appendix \ref{ident}.

Some of the loop diagrams discussed above are UV divergent and require
renormalization. This results in the subtraction of the perturbative part
which is uninteresting from the point of view of the hadronic component of the
photon. To illustrate this problem consider loop integral (\ref{loop2}) with
$\mathcal{N}=1$. Performing $d^{D}\kappa$ integration gives
\begin{align}
\mathcal{J} &  =\frac{i}{16\pi^{2}P^{+}}\left(  \frac{4\pi e^{-\gamma}%
}{\Lambda^{2}}\right)  ^{\epsilon}\frac{1}{\epsilon}%
{\displaystyle\sum\limits_{i,j=1}^{4n+1}}
f_{i}\eta_{i}^{4n}\,f_{j}\eta_{j}^{4n}\left[  1-\bar{u}\eta_{i}+u\eta
_{j}+u\bar{u}\,p^{2}\right]  ^{-\epsilon}\nonumber\\
&  =\frac{i}{16\pi^{2}P^{+}}\left(  \frac{4\pi e^{-\gamma}}{\Lambda^{2}%
}\right)  ^{\epsilon}%
{\displaystyle\sum\limits_{i,j=1}^{4n+1}}
f_{i}\eta_{i}^{4n}\,f_{j}\eta_{j}^{4n}\left(  \frac{1}{\epsilon}-\ln\left[
1-\bar{u}\eta_{i}+u\eta_{j}+u\bar{u}\,p^{2}\right]  \right)  \label{I1}%
\end{align}
with $\bar{u}=u-1$. Because of (\ref{eq:ident}) the coefficient of the
$1/\epsilon$ pole is equal to $1$. For $\mathcal{N}$ involving negative powers
of $\eta_{i}$ (like $\mathcal{N}=M_{k}$ for example) the coefficient of the
pole is $0$ and no subtraction is needed. For constant mass ($n=0$)
$G(z)=z+1-r^{2}$ and for zero mass (current masses are zero in the chiral
limit) there is only one solution of $G(z)=0$, namely $\eta_{1}=-1$. Hence the
perturbative part of the loop integral (\ref{loop2}) reads%
\begin{equation}
\mathcal{J}_{\text{pert}}=\frac{i}{16\pi^{2}P^{+}}\left(  \frac{4\pi
e^{-\gamma}}{\Lambda^{2}}\right)  ^{\epsilon}\left(  \frac{1}{\epsilon}%
-\ln\left[  u\bar{u}\,p^{2}\right]  \right)  .\label{Ipert}%
\end{equation}
This result can be of course obtained by standard techniques for $M=0$.
Renormalization in the $\overline{\text{MS}}$ scheme proceeds by subtracting
the pole only. Here we subtract full perturbative contribution and go back to
$D=4$ ($\epsilon=0$) dimensions which gives%
\begin{equation}
\mathcal{J}_{\text{sub}}=-\frac{i}{16\pi^{2}P^{+}}%
{\displaystyle\sum\limits_{i,j=1}^{4n+1}}
f_{i}\eta_{i}^{4n}\,f_{j}\eta_{j}^{4n}\ln\left[  \frac{1-\bar{u}\eta_{i}%
+u\eta_{j}+u\bar{u}\,p^{2}}{u\bar{u}\,p^{2}}\right]  \label{Isub}%
\end{equation}
where we have used again (\ref{eq:ident}). Note that subtraction occurs only
for terms which do not involve $M_{k}$ or $M_{k-P}$, and these terms are
always UV divergent. In other words in the chiral limit all perturbative
photon DA are either UV divergent or identically zero.

\section{Photon DAs in nonlocal model}

\label{photo}

Before we proceed with photon DAs and systematically present our results we
have to fix numerical constants appearing in the definitions
(\ref{eq:def_tensor})--(\ref{eq:def_axial}). We have already introduced the
expression for pion decay constant (\ref{eq:Fpi}) which is used to fix model
parameters. Next we consider the quark condensate given as the trace of the
quark propagator which reads in Euclidean metric%
\begin{equation}
\left\langle \bar{\psi}\psi\right\rangle =-\frac{N_{c}}{4\pi^{2}}\int
dk_{\text{E}}^{2}\frac{k_{\text{E}}^{2}\,M\left(  k_{\text{E}}^{2}\right)
}{k_{\text{E}}^{2}+M^{2}\left(  k_{\text{E}}^{2}\right)  },
\label{eq:condensate}%
\end{equation}
which in our model turns out to be simply%
\begin{equation}
\left\langle \bar{\psi}\psi\right\rangle =-\frac{N_{c}M\Lambda^{2}}{4\pi^{2}%
}\sum_{i=1}^{4n+1}f_{i}\eta_{i}^{2n}\left(  1+\eta_{i}\right)  \,\ln\left(
1+\eta_{i}\right)  . \label{eq:condensate1}%
\end{equation}
Numerical values obtained from this formula coincide with those of
ref. ~\cite{Praszalowicz:2001wy} if we use model parameters corresponding to
$F_{\pi}$ obtained from the Pagels-Stokar formula \cite{Pagels:1979hd}.

The formula for magnetic susceptibility $\chi_{m}$ in the nonlocal model used
in ref. ~\cite{Dorokhov:2006qm} :%
\begin{equation}
\chi_{m}=\frac{N_{c}}{4\pi^{2}\left\langle \bar{\psi}\psi\right\rangle }\int
dk_{\text{E}}^{2}\frac{k_{\text{E}}^{2}\,\left(  M\left(  k_{\text{E}}%
^{2}\right)  -k_{\text{E}}^{2}M^{\,\prime}\left(  k_{\text{E}}^{2}\right)
\right)  }{\left(  k_{\text{E}}^{2}+M^{2}\left(  k_{\text{E}}^{2}\right)
\right)  ^{2}} \label{chim}%
\end{equation}
(with
$M^{\prime}(k^2_{\mathrm{E}})=
dM(k_{\mathrm{E}})/dk^2_{\mathrm{E}}$)
reduces in our case to%
\begin{align}
\chi_{\mathrm{m}}  &  =\frac{N_{c}M}{4\pi^{2}\left\langle \bar{q}%
q\right\rangle }\,\sum_{i,j=1}^{4n+1}f_{i}f_{j}\eta_{i}^{4n}\left(  1+\eta
_{j}\right)  \left[  \eta_{j}^{2n}+2n\left(  1+\eta_{j}\right)  \eta
_{j}^{2n-1}\right] \label{chim-ours}\\
&  \qquad\left\{  \frac{\varepsilon_{ij}}{\eta_{i}-\eta_{j}}\left(
\log\left(  1+\eta_{i}\right)  -\log\left(  1+\eta_{j}\right)  \right)
+\frac{\delta_{ij}}{1+\eta_{i}}\right\} \nonumber
\end{align}
where $\epsilon_{ij}$ is $0$ for $i=j$ and $1$ otherwise, while $\delta_{ij}$
is Kronecker delta. Numerical values of $\left\langle \bar{\psi}%
\psi\right\rangle $ and $\chi_{m}$ for the present set of model parameters are
listed in Table \ref{tab:1}. Note that in fact we do not have to use
(\ref{chim-ours}) to calculate $\chi_{m}$ since it can be retrieved from the
normalization condition of $\phi_{T}(u)$. Numerical values of $\chi_{m}$
obtained both in ways agree proving consistency of our calculations and
definitions (\ref{eq:def_tensor}).

\begin{table}[ptb]
\begin{centering}
\begin{tabular}{|c|c|c|}
\hline
\multicolumn{3}{|c|}{$M=300\,\mathrm{MeV}$}\tabularnewline
\hline
\hline
$n=1$ & $\left\langle \bar{\psi}\psi\right\rangle =
-\left(277\,\mathrm{MeV}\right)^{3}$ & $\chi_{m}=2.30\,\mathrm{GeV}^{-2}$\tabularnewline
\hline
$n=5$ & $\left\langle \bar{\psi}\psi\right\rangle =
-\left(230\,\mathrm{MeV}\right)^{3}$ & $\chi_{m}=3.75\,\mathrm{GeV}^{-2}$\tabularnewline
\hline
\hline
\multicolumn{3}{|c|}{$M=350\,\mathrm{MeV}$}\tabularnewline
\hline
\hline
$n=1$ & $\left\langle \bar{\psi}\psi\right\rangle =
-\left(253\,\mathrm{MeV}\right)^{3}$ & $\chi_{m}=2.85\,\mathrm{GeV}^{-2}$\tabularnewline
\hline
$n=5$ & $\left\langle \bar{\psi}\psi\right\rangle =
-\left(208\,\mathrm{MeV}\right)^{3}$ & $\chi_{m}=4.71\,\mathrm{GeV}^{-2}$\tabularnewline
\hline
\hline
\multicolumn{3}{|c|}{$M=400\,\mathrm{MeV}$}\tabularnewline
\hline
\hline
$n=1$ & $\left\langle \bar{\psi}\psi\right\rangle =
-\left(236\,\mathrm{MeV}\right)^{3}$ & $\chi_{m}=3.34\,\mathrm{GeV}^{-2}$\tabularnewline
\hline
$n=5$ & $\left\langle \bar{\psi}\psi\right\rangle =
-\left(192\,\mathrm{MeV}\right)^{3}$ & $\chi_{m}=5.58\,\mathrm{GeV}^{-2}$\tabularnewline
\hline
\end{tabular}
\par\end{centering}
\caption{Numerical values of the quark condensate $\left\langle \bar{\psi}%
\psi\right\rangle $ obtained using model parameters from Tab.\ref{tab:0} and
magnetic susceptibility $\chi_{m}$ used in the calculations.}%
\label{tab:1}%
\end{table}

To calculate $f_{3\gamma}$ we have used Euclidean formula from
ref. \cite{Dorokhov:2006qm}:%

\begin{equation}
f_{3\gamma}=-\frac{N_{c}}{4\pi^{2}}\,\int dk_{\text{E}}^{2}\frac
{\,M^{2}\left(  k_{\text{E}}^{2}\right)  }{k_{\text{E}}^{2}+M^{2}\left(
k_{\text{E}}^{2}\right)  },
\end{equation}
which in our model transforms into%
\begin{equation}
f_{3\gamma}=\frac{N_{c}M^{2}}{4\pi^{2}}\,\sum_{i=1}^{4n+1}f_{i}\,\ln\left(
1+\eta_{i}\right)  . \label{f3g2}%
\end{equation}
Numerical values of $f_{3\gamma}$ are listed in Table \ref{tab:f3}.

\begin{table}[ptb]
\begin{centering}
\begin{tabular}{|c|c|}
\hline
\multicolumn{2}{|c|}{$M=300\,\mathrm{MeV}$}\tabularnewline
\hline
\hline
$n=1$ & $f_{3\gamma}=-0.0095\,\mathrm{GeV}^{2}$\tabularnewline
\hline
$n=5$ & $f_{3\gamma}=-0.0093\,\mathrm{GeV}^{2}$\tabularnewline
\hline
\hline
\multicolumn{2}{|c|}{$M=350\,\mathrm{MeV}$}\tabularnewline
\hline
\hline
$n=1$ & $f_{3\gamma}=-0.0095\,\mathrm{GeV}^{2}$\tabularnewline
\hline
$n=5$ & $f_{3\gamma}=-0.0092\,\mathrm{GeV}^{2}$\tabularnewline
\hline
\hline
\multicolumn{2}{|c|}{$M=400\,\mathrm{MeV}$}\tabularnewline
\hline
\hline
$n=1$ & $f_{3\gamma}=-0.0094\,\mathrm{GeV}^{2}$\tabularnewline
\hline
$n=5$ & $f_{3\gamma}=-0.0091\,\mathrm{GeV}^{2}$\tabularnewline
\hline
\end{tabular}
\par\end{centering}
\caption{Numerical values of $f_{3\gamma}$ obtained using model parameters
from Tab.\ref{tab:0}.}%
\label{tab:f3}%
\end{table}

Phenomenological values of $\left\langle \bar{\psi}\psi\right\rangle $,
$\chi_{m}$ and $f_{3\gamma}$ are well known only for the quark condensate:
approximately $-(250$ MeV$)^{3}$ \cite{Shifman:1978by} at low momentum scale.
This value is still used in more recent phenomenological applications
\cite{Ball:2002ps}. Magnetic susceptibility is still a subject of large
phenomenological uncertainties. Different estimates are nicely summarized in
ref. ~\cite{Pimikov:2008ay} where it is shown that $\chi_{m}\simeq2.5\div5.5 $
GeV$^{-2}$ with some preference to the values around 4.3 GeV$^{-2}$. Finally
the value of $f_{3\gamma}$ obtained in different low energy models, as
discussed in ref. ~\cite{Dorokhov:2006qm}, is negative and of the order of
$-0.004$ GeV$^{2}$. Our values are here factor of 2 smaller ($\sim0.0094$
GeV$^{2}$), however they are almost insensitive to actual model parameters. On
the other hand magnetic susceptibility is quite sensitive to $M$ and $n$ (see
eqs. (\ref{eq:massF}) and (\ref{Fkdef})) remaining, however, within the range
of acceptable phenomenological values discussed in ref. ~\cite{Pimikov:2008ay}.
Similarly $\left\langle \bar{\psi}\psi\right\rangle $ varies with $M$ and $n$,
however, for the preferred value of the constituent quark mass $M=350$ MeV it
is quite close to the phenomenological estimates. From this point of view our
model satisfactorily describes low energy observables relevant for photon DAs.

\subsection{Leading twist distributions}

\label{leadi}

\subsubsection{Tensor photon DA}

\label{tenso}

Tensor twist-2 amplitude has been already discussed in
refs. ~\cite{Petrov:1998kg} and also \cite{Praszalowicz:2001wy}
in a model with
the local vertex only. Here we extend discussion to off-shell photons and
calculate the correction appearing due to the modified vertex
\eqref{eq:vertex}. In the case of leading twist tensor DA we obtain the
following expression using local current:%
\begin{align}
\phi_{T}^{\left(  0\right)  }\left(  u,P^{2}\right)   &  =\frac{i4N_{c}P^{+}%
}{\left\langle \bar{\psi}\psi\right\rangle \chi_{m}F_{T}\left(  P^{2}\right)
}\int\frac{d^{D}k}{\left(  2\pi\right)  ^{D}}\,\delta\left(  k\cdot
n-u\,n\cdot P\right) \nonumber\\
&  \frac{\bar{u}\,M_{k}-u\,M_{k-P}}{\left(  k^{2}-M_{k}^{2}+i\epsilon\right)
(\left(  k-P\right)  ^{2}-M_{k-P}^{2}+i\epsilon)}, \label{eq:tensor_tw2_1}%
\end{align}
where $\bar{u}=u-1.$ Notice, that special choice of the contour described in
\cite{Praszalowicz:2001wy} allows for passing to Euclidean space.
Therefore we can use Schwinger representation for scalar propagators and proceed in the spirit of \cite{Dorokhov:2006qm}. The result reads:%
\begin{equation}
\phi_{T}^{(0)}\left(  u,P^{2}\right)  =\frac{-N_{c}M}{4\pi^{2}\left\langle
\bar{\psi}\psi\right\rangle \chi_{m}F_{T}\left(  P^{2}\right)  }\sum
_{i,j=1}^{4n+1}f_{i}f_{j}\left(  \bar{u}\eta_{i}^{2n}\eta_{j}^{4n}-u\eta
_{i}^{4n}\eta_{j}^{2n}\right)  \ln\left(  1+u\bar{u}p^{2}-\bar{u}\eta
_{i}+u\eta_{j}\right)  . \label{eq:tensor_tw2-6}%
\end{equation}
for $0\leq u\leq1$.

\begin{figure}[ptb]
\begin{centering}
\begin{tabular}{ll}
a) & b)\tabularnewline
\includegraphics[width=8cm]{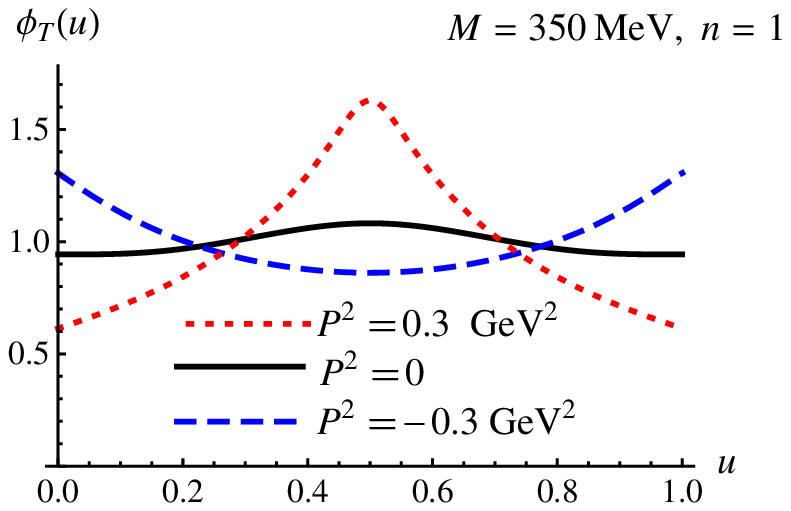} & \includegraphics[width=8cm]{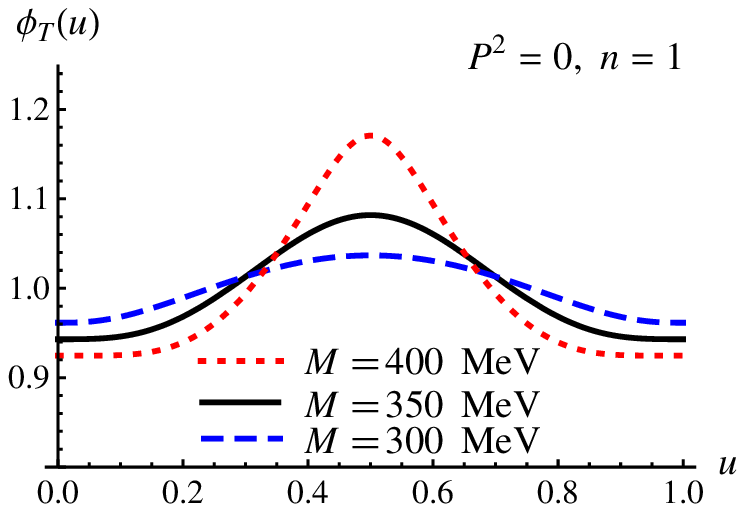}\tabularnewline
c) & d)\tabularnewline
\includegraphics[width=8cm]{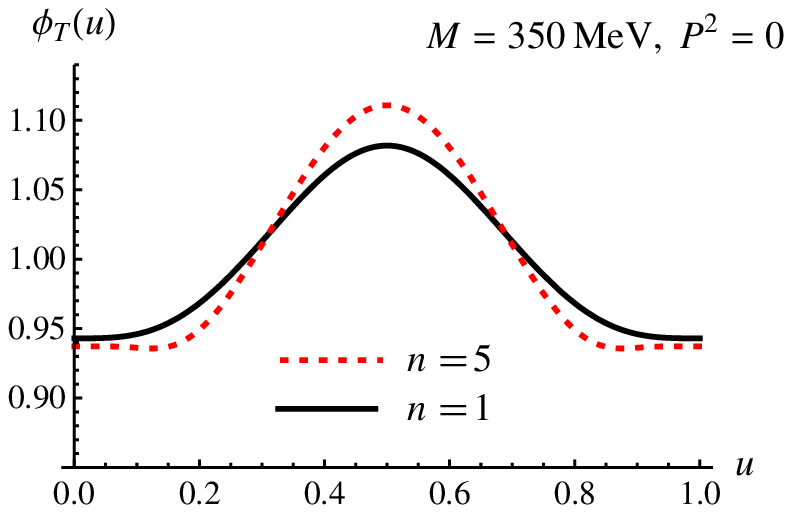} & \includegraphics[width=8cm]{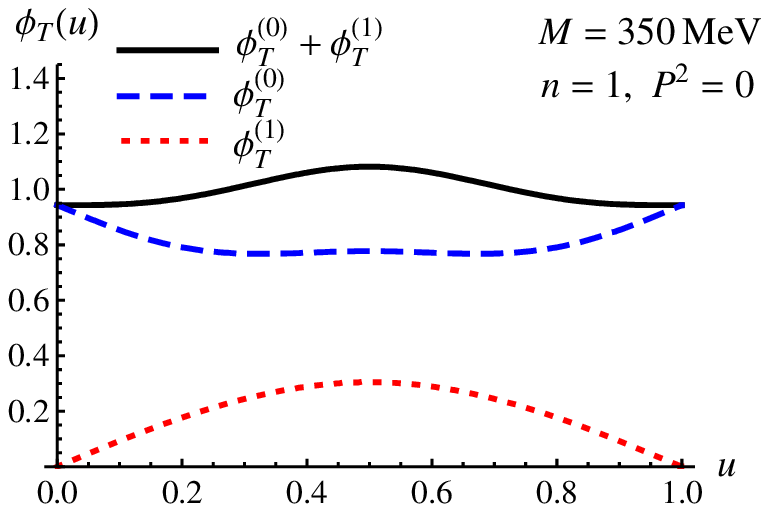}\tabularnewline
\end{tabular}
\par\end{centering}
\caption{Leading twist tensor photon DA for:
a) different photon virtualities and fixed $M=350$~MeV and $n=1$,
b) different $M$ and fixed $n=1$ and $P^2=0$,
c) different $n$ and fixed $M=350$~MeV and $P^2=0$,
d) decomposition into contributions corresponding to local (dashed)
and non-local (dotted) parts of the vector vertex
for $M=350$~MeV, $n=1$ and $P^2=0$.}%
\label{fig:tensor}%
\end{figure}

Now let us consider the part coming from the nonlocal part of the vertex
\eqref{eq:vertex}. It is given by the integral%
\begin{align}
\phi_{T}^{\left(  1\right)  }\left(  u,P^{2}\right)   &  =\frac{-i8N_{c}%
\,P^{+}}{\left\langle \bar{\psi}\psi\right\rangle \chi_{m}F_{T}\left(
P^{2}\right)  }\int\frac{d^{D}k}{\left(  2\pi\right)  ^{D}}\,\delta\left(
k\cdot n-u\,n\cdot P\right)  \label{eq:eq:tensor_tw2_7}\\
&  \,\frac{\left(  M_{k}-\,M_{k-P}\right)  \left(  \varepsilon_{\bot}\cdot
k_{\bot}\right)  ^{2}}{\left(  k^{2}-M_{k}^{2}+i\epsilon\right)  (\left(
k-P\right)  ^{2}-M_{k-P}^{2}+i\epsilon)\left(  2k\cdot P-P^{2}+i\mu\right)
},\nonumber
\end{align}
Notice that there appears additional denominator $2k\cdot P-P^{2}+i\mu$, where
$+i\mu$ prescription introduced at this stage is completely arbitrary.
However, as already explained in sect. \ref{nonlo} the residue of this pole is
zero so that it does not contribute to the amplitude irrespectively of the
sign of $\mu$. Therefore in the following we shall always omit contribution of
this spurious pole. After performing the integration as described in appendix
\ref{light} we finally obtain%
\begin{align}
\phi_{T}^{\left(  1\right)  }\left(  u,P^{2}\right)   &  =\,\frac{MN_{c}%
}{4\pi^{2}\left\langle \bar{\psi}\psi\right\rangle \chi_{m}F_{T}\left(
P^{2}\right)  }\,\sum_{i,j}^{4n+1}f_{i}f_{j}\frac{\eta_{i}^{2n}\eta_{j}%
^{2n}\left(  \eta_{i}^{2n}-\eta_{j}^{2n}\right)  }{\eta_{i}-\eta_{j}%
}\label{eq:tensor_tw2_13}\\
&  \left(  1+u\bar{u}p^{2}-\bar{u}\eta_{i}+u\eta_{j}\right)  \ln\left(
1+u\bar{u}p^{2}-\bar{u}\eta_{i}+u\eta_{j}\right)  \nonumber
\end{align}
for $0\leq u\leq1$. The full twist-2 tensor photon DA is given by%
\begin{equation}
\phi_{T}=\phi_{T}^{\left(  0\right)  }+\phi_{T}^{\left(  1\right)
}.\label{eq:tensor_tw2_14}%
\end{equation}
We plot this function in fig. \ref{fig:tensor} for several values of photon
virtuality and constituent quark mass. Notice that the nonlocal part of the
quark-photon vertex is small and the full amplitude is almost equal to the
local one. The resulting DA is almost flat for real photons and does not
vanish at the end points.

\begin{figure}[ptb]
\begin{centering}
\begin{tabular}{ll}
a) & b)\tabularnewline
\includegraphics[width=8cm]{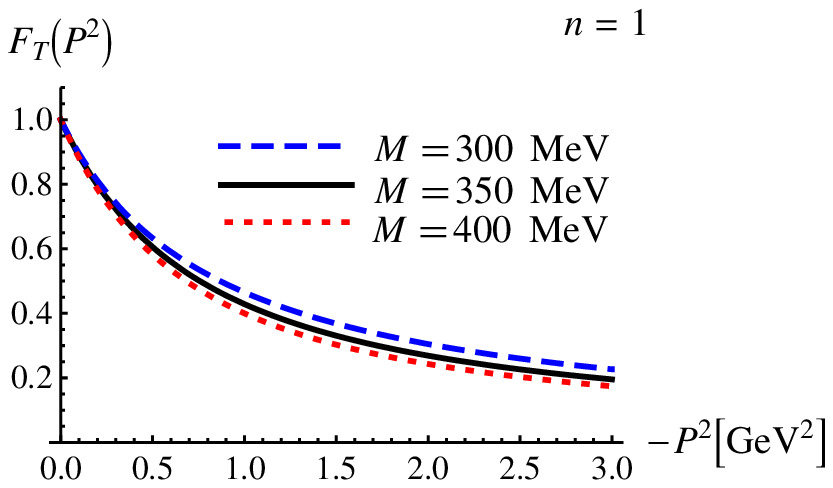} & \includegraphics[width=8cm]{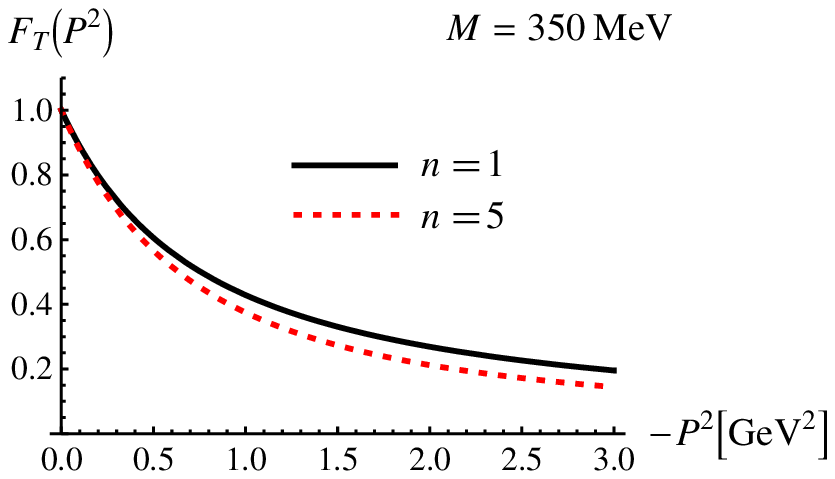}\tabularnewline
\end{tabular}\caption{Tensor form factor for:
a) fixed $n=1$ and different $M$,
b) fixed $M=350$~MeV and two different $n=1,5$.}
\end{centering}
\label{fig:tensor_ff}
\end{figure}

Tensor form factor is shown in fig. \ref{fig:tensor_ff}. It can be in principle calculated by
analytical integration, which has to be performed carefully because of the
complex numbers under logarithms.

\subsubsection{Vector photon DA}

\label{vecto}

Calculation of vector twist-2 amplitude proceeds in a similar way. After
performing the traces we get%
\begin{equation}
\psi_{V}(u)=\frac{-i4P^{+}N_{c}}{f_{3\gamma}F_{V}(P^{2})\varepsilon^{+}}%
{\displaystyle\int}
\frac{d^{D}k}{(2\pi)^{D}}\frac{\left(  T_{V}^{(0)}+T_{V}^{(1)}\right)
\,\delta\left(  k\cdot n-u\,n\cdot P\right)  }{(k^{2}-M_{k}^{2}+i\epsilon
)((k-P)^{2}-M_{k-P}^{2}+i\epsilon)}.
\end{equation}
where $T_V^{(0)}$ and $T_V^{(1)}$ stand for traces corresponding to local and
nonlocal parts of the photon vertex respectively:%
\begin{align}
T_V^{(0)}  &  =\varepsilon^{+}(M_{k}M_{k-P}+\vec{k}_{\bot}^{2}-P^{2}u\bar
{u})-(\vec{\varepsilon}_{\bot}\cdot\vec{k}_{\bot})P^{+}(\bar{u}+u),\nonumber\\
T_V^{(1)}  &  =-\frac{\left(  M_{k}-M_{k-P}\right)  \left(  \bar{u}%
M_{k}+uM_{k-P}\right)  }{k^{-}+\bar{u}\frac{P^{2}}{P^{+}}}\left[
\varepsilon^{+}\left(  k^{-}-u\frac{P^{2}}{P^{+}}\right)  -2\,(\vec
{\varepsilon}_{\bot}\cdot\vec{k}_{\bot})\right]  .
\end{align}
Note that single powers of $(\vec{\varepsilon}_{\bot}\cdot\vec{k}_{\bot})$
integrate to zero.

In the case of twist 2 vector DA we have to subtract the perturbative piece
corresponding to $M\left(  k\right)  =0$. Then, the contribution to the vector
photon DA coming from the local part of the vertex consists of two parts%
\begin{align}
\phi_{V}^{(0,a)}\left(  u,P^{2}\right)   &  =\frac{N_{c}}{4\pi^{2}f_{3\gamma
}F_{V}(P^{2})}\sum_{i,j=1}^{4n+1}f_{i}f_{j}\left(  \Lambda^{2}\eta_{i}%
^{4n}\eta_{j}^{4n}\left(  \bar{u}\eta_{i}-u\eta_{j}-1\right)  +M^{2}\eta
_{i}^{2n}\eta_{j}^{2n}\right) \nonumber\\
&  \qquad\qquad\qquad\qquad\qquad\qquad\qquad\qquad\qquad\ln\left(  1+u\bar
{u}p^{2}-\bar{u}\eta_{i}+u\eta_{j}\right)  . \label{eq:V_tw2_0a}%
\end{align}
and%
\begin{equation}
\phi_{V}^{(0,b)}\left(  u,P^{2}\right)  =\frac{N_{c}}{4\pi^{2}f_{3\gamma}%
F_{V}(P^{2})}\left(  -2u\bar{u}P^{2}\right)  \sum_{i,j=1}^{4n+1}f_{i}f_{j}%
\eta_{i}^{4n}\eta_{j}^{4n}\ln\left(  \frac{1+u\bar{u}p^{2}-\bar{u}\eta
_{i}+u\eta_{j}}{u\bar{u}p^{2}}\right)  . \label{eq:V_tw2_0b}%
\end{equation}
The addition coming from the nonlocal part of the current can be conveniently
split into a sum of two contributions%
\begin{align}
\phi_{V}^{(1,a)}(u)  &  =\frac{-N_{c}}{4\pi^{2}f_{3\gamma}F_{V}(P^{2})}%
\,M^{2}\sum_{i,j=1}^{4n+1}f_{i}f_{j}\left(  \eta_{j}^{2n}-\eta_{i}%
^{2n}\right)  \left(  \bar{u}\eta_{j}^{2n}+u\eta_{i}^{2n}\right) \nonumber\\
&  \qquad\qquad\qquad\qquad\qquad\qquad\qquad\qquad\qquad\ln\left(  1\frac{{}%
}{{}}+u\bar{u}p^{2}-\bar{u}\eta_{i}+u\eta_{j}\right)  ,\label{eq:V_tw2_1a}\\
& \nonumber\\
\phi_{V}^{(1,b)}(u)  &  =\frac{N_{c}}{4\pi^{2}f_{3\gamma}F_{V}(P^{2}%
)}(1-2u)\frac{M^{2}P^{2}}{\Lambda^{2}}\sum_{i,j=1}^{4n+1}f_{i}f_{j}%
\frac{\left(  \eta_{i}^{2n}-\eta_{j}^{2n}\right)  \left(  \bar{u}\eta_{j}%
^{2n}+u\eta_{i}^{2n}\right)  }{\left(  \eta_{i}-\eta_{j}\right)
}\nonumber\\
&  \qquad\qquad\qquad\qquad\qquad\qquad\qquad\qquad\qquad\ln\left(  1\frac{{}%
}{{}}+u\bar{u}p^{2}-\bar{u}\eta_{i}+u\eta_{j}\right)  . \label{eq:V_tw2_1b}%
\end{align}
Notice that subtraction concerns only $\phi_{V}^{\mathrm{\left(  0,b\right)
}}$ part since $\phi_{V}^{\mathrm{\left(  1\right)  }}$ is always proportional
to mass.

\begin{figure}[ptb]
\begin{centering}
\begin{tabular}{ll}
a) & b)\tabularnewline
\includegraphics[width=8cm]{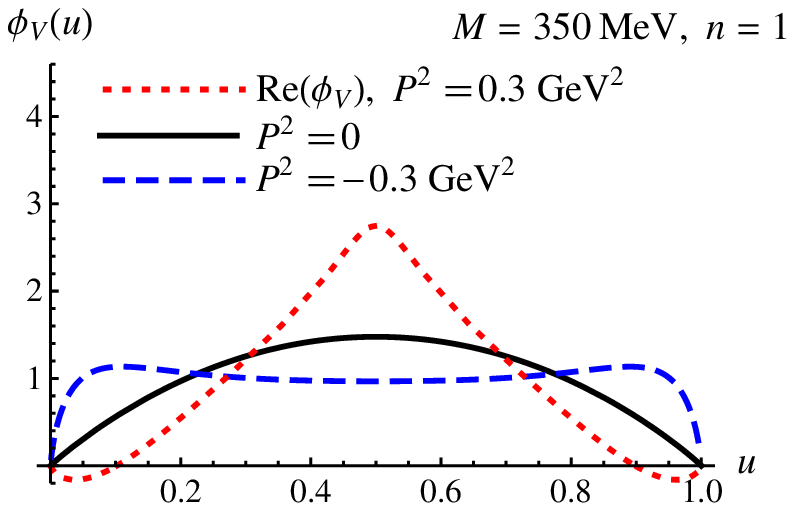} & \includegraphics[width=8cm]{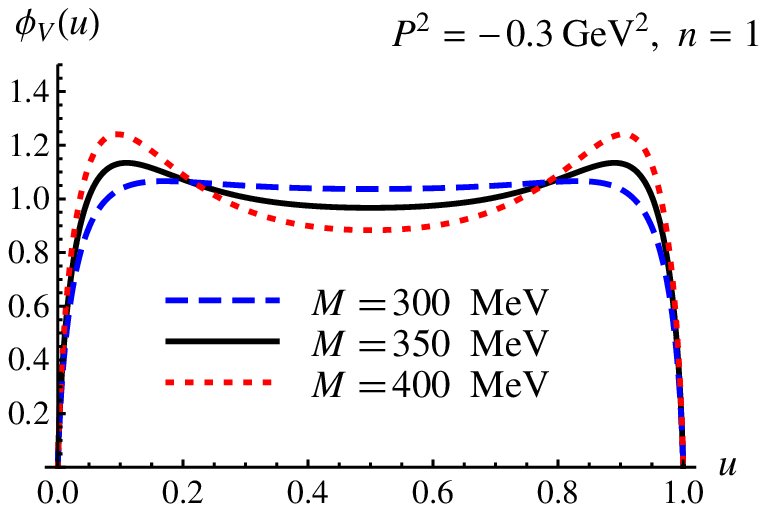}\tabularnewline
c) & d)\tabularnewline
\includegraphics[width=8cm]{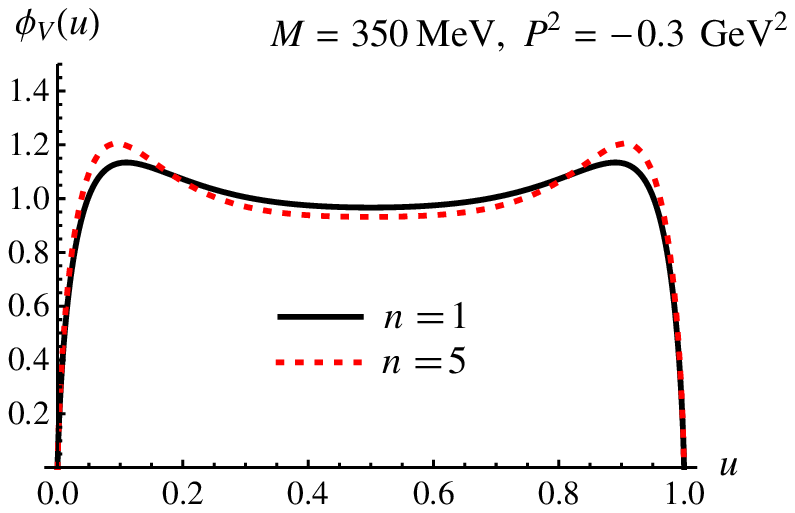} & \includegraphics[width=8cm]{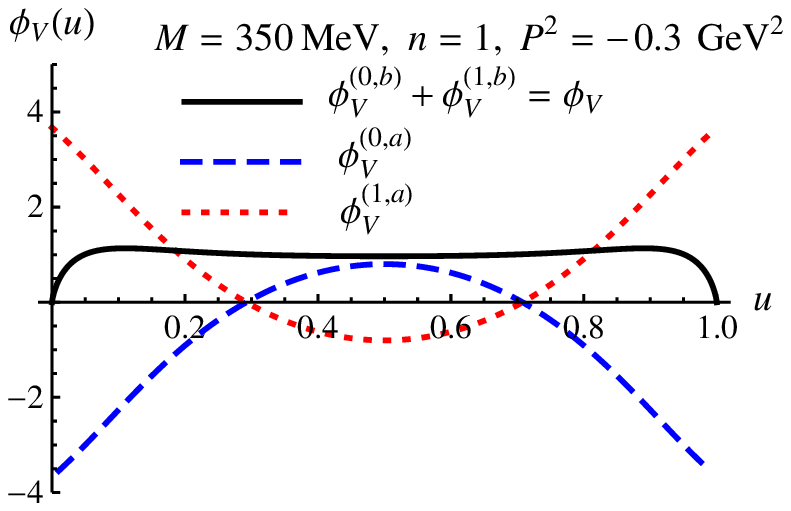}\tabularnewline
\end{tabular}
\end{centering}
\caption{Leading twist vector photon DA for:
a) different photon virtualities and fixed $M=350$~MeV and $n=1$,
b) different $M$ and fixed $n=1$ and $P^2=-0.3$~GeV$^2$,
c) different $n$ and fixed $M=350$~MeV and $P^2=-0.3$~GeV$^2$,
d) decomposition into different contributions, as described in the main text; notice the exact cancelation of $\phi_V^{(0,a)}$ and $\phi_V^{(1,a)}$
following from the gauge invariance. For positive $P^2$ we show only real part of DA.}%
\label{fig:vector}%
\end{figure}

In the tensor case the only effect due to the local vertex is a small change
in the shape of the distribution. The situation is different for vector DA.
When we use local current only, the vector distribution alternates
 in sign (recall that leading twist DAs have probabilistic interpretation).
Only when we include the nonlocal part of the vertex, contributions $\phi
_{V}^{(0,a)}$ and $\phi_{V}^{(1,a)}$ cancel exactly for any $P^{2}$. This is
explicitly shown in fig.\ref{fig:vector}.d.  As a consequence $\phi_{V}$
is effectively the sum
of $\phi_{V}^{(0,b)}$ and $\phi_{V}^{(1,b)}$. We plot this function in fig.
\ref{fig:vector} for different sets of model parameters.

Furthermore, since both $\phi_{V}^{(0,b)}$ and $\phi_{V}^{(1,b)}$ are
explicitly proportional to $P^{2}$, normalization conditions
(\ref{eq:normaliz2}) require that $F_{V}(0)=0$ as it should be in accordance
with the conservation of the vector current. This condition would be violated
if not for the nonlocal part of the photon vertex.

\begin{figure}[ptb]
\begin{centering}
\begin{tabular}{ll}
a) & b)\tabularnewline
\includegraphics[width=8cm]{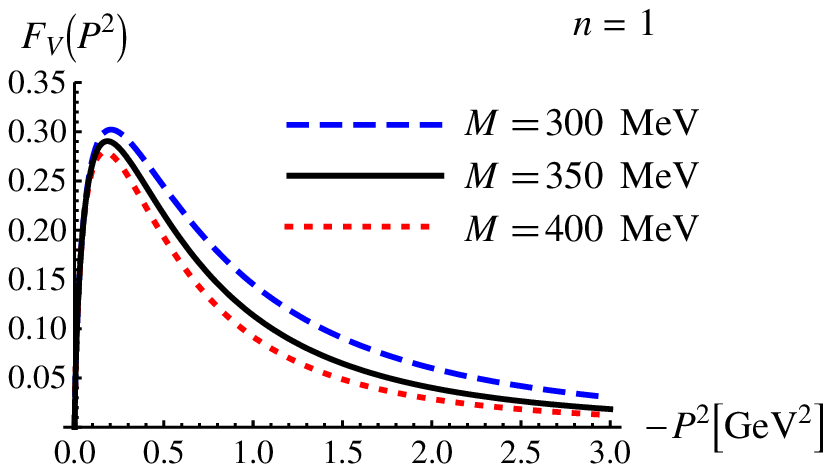} & \includegraphics[width=8cm]{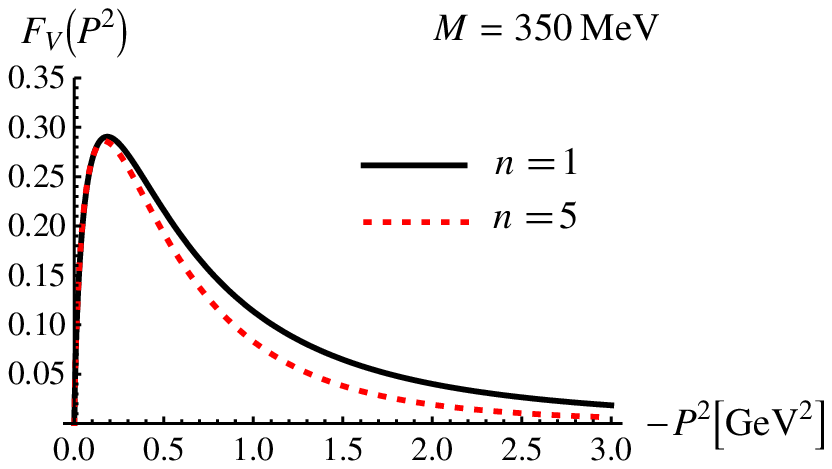}\tabularnewline
\end{tabular}\caption{Vector form factor for:
a) fixed $n=1$ and different $M$,
b) fixed $M=350$~MeV and two different choices of $n=1,5$.
Notice that the form factor vanishes for
zero virtuality as required by vector current conservation.}
\end{centering}
\label{fig:vector_formf}
\end{figure}

\subsection{Higher twist distributions}

\label{highe}

For higher twist distributions we encounter an additional difficulty. As
shown  in appendix \ref{light} it turns out that they are in fact generalized
functions. Due to $k^{-}=k\cdot\tilde{n}$ occurring in the numerator
($k^{+}=k\cdot n$ is fixed -- see delta function in \eqref{eq:tensor_tw2_1}),
additional end point delta functions appear. They are crucial for Lorentz
invariance of the integrals and in consequence for the correct normalization
of the distributions. These singularities were already discussed in
ref. ~\cite{Dorokhov:2006qm}.

\subsubsection{Tensor DAs}

\label{tensoh}

We start with tensor DAs. For twist 3 tensor amplitude we have%
\begin{equation}
\psi_{T}(u)=\frac{4N_{c}}{\left\langle \bar{\psi}\psi\right\rangle
F_{T}(P^{2})}\frac{P^{+}}{\varepsilon^{+}}%
{\displaystyle\int}
\frac{d^{4}k}{(2\pi)^{4}}\frac{\left(  T_{T}^{(0)}+T_{T}^{(1)}\right)
\,\delta\left(  k\cdot n-uP^{+}\right)  }{(k^{2}-M_{k}^{2}+i\epsilon
)((k-P)^{2}-M_{k-P}^{2}+i\epsilon)}%
\end{equation}
where $T_{T}^{(0)}$ and $T_{T}^{(1)}$ stand for traces corresponding to local
and nonlocal parts of the photon vertex respectively:%
\begin{align}
T_{T}^{(0)} &  =-\frac{i}{2}\varepsilon^{+}\left[  P^{+}(M_{k}-M_{k-P})\left(
k^{-}+\bar{u}\frac{P^{2}}{P^{+}}\right)  -P^{2}(M_{k}+M_{k-P})\right]
,\label{tens_3_loc}\\
T_{T}^{(1)} &  =\frac{i}{2}\frac{P^{+}\left(  M_{k}-M_{k-P}\right)  }%
{k^{-}+\bar{u}\frac{P^{2}}{P^{+}}}\left(  k^{-}-u\frac{P^{2}}{P^{+}}\right)
\left[  \varepsilon^{+}\left(  k^{-}-u\frac{P^{2}}{P^{+}}\right)  -2\vec
{k}_{\bot}\cdot\vec{\varepsilon}_{\bot}\right]  \,.\label{tens_3_nl}%
\end{align}
Note that even powers of $\vec{k}_{\bot}\cdot\vec{\varepsilon}_{\bot}$
integrate to zero under $d^{2}\vec{k}_{\bot}$.

Luckily in the case of $\psi_{T}(u)$ contributions involving $k^{-}$ cancel
out in the sum of (\ref{tens_3_loc}) and (\ref{tens_3_nl}) and the final
result is the sum of two pieces%
\begin{align}
\psi_{T}^{(a)}(u)  &  =\frac{N_{c}M}{8\pi^{2}\left\langle \bar{\psi}%
\psi\right\rangle \chi_{m}F_{T}(P^{2})}\left(  \chi_{m}P^{2}\right) \\
&  \sum_{i,j=1}^{4n+1}f_{i}f_{j}\eta_{i}^{2n}\eta_{j}^{2n}\left[  (\eta
_{j}^{2n}+\eta_{i}^{2n})+2\left(  1-2u\right)  \left(  \eta_{j}^{2n}-\eta
_{i}^{2n}\right)  \right]  \ln\left(  1\frac{{}}{{}}+u\bar{u}p^{2}-\bar{u}%
\eta_{i}+u\eta_{j}\right) \nonumber
\end{align}
and%
\begin{align}
\psi_{T}^{(b)}(u)  &  =\frac{-N_{c}M}{8\pi^{2}\left\langle \bar{\psi}%
\psi\right\rangle \chi_{m}F_{T}(P^{2})}\left(  \chi_{m}P^{2}\right)
\frac{P^{2}}{\Lambda^{2}}\left(  1-2u\right)  ^{2}\nonumber\\
&  \sum_{i,j=1}^{4n+1}f_{i}f_{j}\eta_{i}^{2n}\eta_{j}^{2n}\frac{\eta_{j}%
^{2n}-\eta_{i}^{2n}}{\eta_{j}-\eta_{i}}\ln\left(  1\frac{{}}{{}}+u\bar{u}%
p^{2}-\bar{u}\eta_{i}+u\eta_{j}\right)  .
\end{align}
Note that $\psi_{T}$ is proportional to the same normalization constant as
$\phi_{T}$ times $\left(  \chi_{m}P^{2}\right)  $ which means that it
decouples for real photons.

In the case of twist $4$ tensor amplitude the $\delta$ function contributions
do not cancel out. Performing Dirac traces we have%
\begin{equation}
h_{T}(u)=\frac{4N_{c}P^{+\,2}}{\left\langle \bar{\psi}\psi\right\rangle
\,F_{T}(P^{2})}%
{\displaystyle\int}
\frac{d^{4}k}{(2\pi)^{4}}\frac{\left(  R_{T}^{(0)}+R_{T}^{(1)}\right)
\,\delta\left(  k\cdot n-u\,n\cdot P\right)  }{(k^{2}-M_{k}^{2}+i\epsilon
)((k-P)^{2}-M_{k-P}^{2}+i\epsilon)}%
\end{equation}
where again the contributions of local and nonlocal parts of the vector
current have been singled out:%
\begin{align}
R_{T}^{(0)} &  =-i\frac{P^{2}}{P^{+\,2}}\varepsilon^{+}(M_{k}-M_{k-P})\left(
\vec{\varepsilon}_{\bot}\cdot\vec{k}_{\bot}\right)  -i\left[  (M_{k}%
-M_{k-P})k^{-}-M_{k}\frac{P^{2}}{P^{+}}\right]  ,\label{T03}\\
R_{T}^{(1)} &  =\,-i\frac{P^{2}}{P^{+\,2}}\frac{M_{k}-M_{k-P}}{k^{-}+\bar
{u}\frac{P^{2}}{P^{+}}}\,\left(  \vec{\varepsilon}_{\bot}\cdot\vec{k}_{\bot
}\right)  \left[  \varepsilon^{+}\left(  k^{-}-u\frac{P^{2}}{P^{+}}\right)
-2\left(  \vec{\varepsilon}_{\bot}\cdot\vec{k}_{\bot}\right)  \right]
\end{align}
Note that significant simplifications occur since single power of $\vec
{k}_{\bot}\cdot\vec{\varepsilon}$ integrates to zero. Following the steps
described in appendix \ref{light} we finally arrive at the final formula for
$h_{T}(u)$. It is convenient to split it into 4 different pieces -- regular
local vertex contribution:
\begin{align}
h_{T}^{(0,a)}(u) &  =\frac{N_{c}M}{4\pi^{2}\left\langle \bar{\psi}%
\psi\right\rangle \chi_{m}\,F_{T}(P^{2})}\chi_{m}P^{2}%
{\displaystyle\sum\limits_{i,j=1}^{4n+1}}
f_{i}\,f_{j}\eta_{i}^{2n}\eta_{j}^{2n}\left(  u\eta_{i}^{2n}-\bar{u}\eta
_{j}^{2n}\right)  \nonumber\\
&  \quad\quad\quad\quad\quad\quad\quad\quad\quad\quad\quad\quad\quad\ln\left(
1\frac{{}}{{}}+u\bar{u}p^{2}-\bar{u}\eta_{i}+u\eta_{j}\right)  ,\label{ha}%
\end{align}
second local vertex contribution:%
\begin{align}
h_{T\text{ }}^{(0,b)}(u) &  =\frac{-N_{c}M}{4\pi^{2}\left\langle \bar{\psi}%
\psi\right\rangle \chi_{m}\,F_{T}(P^{2})}\chi_{m}\Lambda^{2}%
{\displaystyle\sum\limits_{i,j=1}^{4n+1}}
f_{i}\,f_{j}\,\eta_{i}^{2n}\eta_{j}^{2n}\left(  \eta_{j}^{2n}-\eta_{i}%
^{2n}\right)  \left(  \left(  \eta_{i}-\eta_{j}\right)  +\frac{{}}{{}%
}(1-2u)p^{2}\right)  \nonumber\\
&  \quad\quad\quad\quad\quad\quad\quad\quad\quad\quad\quad\quad\quad\quad
\quad\quad\quad\quad\ln\left(  1\frac{{}}{{}}+u\bar{u}p^{2}-\bar{u}\eta
_{i}+u\eta_{j}\right)  \label{hb}%
\end{align}
that integrates to zero with $\delta$-function contribution
\begin{align}
h_{T\text{ }}^{(0,\mathrm{delta})}(u) &  =\frac{N_{c}M}{4\pi^{2}\left\langle \bar{\psi}%
\psi\right\rangle \chi_{m}\,F_{T}(P^{2})}\chi_{m}\Lambda^{2}%
{\displaystyle\sum\limits_{i,j=1}^{4n+1}}
f_{i}\,f_{j}\,\eta_{i}^{2n}\eta_{j}^{2n}\left(  \eta_{i}^{2n}-\eta_{j}%
^{2n}\right)  \nonumber\\
&  \left[  (1+\eta_{j})\ln\left(  1+\eta_{j}\right)  \delta(u-1)\frac{{}}{{}%
}-(1+\eta_{i})\ln\left(  1+\eta_{i}\right)  \delta(u)\right]  \label{hc}%
\end{align}
and hence does not contribute to the normalization, and the contribution
corresponding to the nonlocal part of the photon vertex:%
\begin{align}
h_{T}^{(1)}(u) &  =\frac{N_{c}M}{4\pi^{2}\left\langle \bar{\psi}%
\psi\right\rangle \,\chi_{m}F_{T}(P^{2})}\chi_{m}P^{2}\,\sum_{i,j=1}%
^{4n+1}f_{i}f_{j}\eta_{i}^{2n}\eta_{j}^{2n}\left(  1\frac{{}}{{}}+u\bar
{u}p^{2}-\bar{u}\eta_{i}+u\eta_{j}\right)  \nonumber\\
&  \qquad\qquad\qquad\qquad\qquad\qquad\qquad\frac{\eta_{i}^{2n}-\eta_{j}%
^{2n}}{\eta_{i}-\eta_{j}}\ln\left(  1\frac{{}}{{}}+u\bar{u}p^{2}-\bar{u}%
\eta_{i}+u\eta_{j}\right)  .\label{h1}%
\end{align}
The delta contribution $h_{T}^{(0,\mathrm{delta})}$ can be rewritten using
the expression (\ref{f3g2}) for $f_{3\gamma}$  and the identities given in
Appendix \ref{ident}:
\begin{align}
h_{T}^{(0,\mathrm{delta})}(u) &  =\frac{-1}
{F_{T}(P^{2})}\left[ \delta(u-1)
+\delta(u)\right]  \label{hcdelt}%
\end{align}
Note that the fact that the sum of (\ref{hb}) and (\ref{hc}) integrates over
$du$ to zero is a consequence of Lorentz invariance discussed in sect.
\ref{loopi} (see eq. (\ref{LorentzB0})). Therefore only $h_{T}^{(0,a)}$ and
$h_{T}^{(1)}(u)$ contribute to the normalization condition (\ref{eq:normaliz1}%
) given by $(\chi_{m}P^{2})$. If not for the $\delta$-term $h_{T}^{(0,b)}$
would also contribute to the norm spoiling the normalization condition.

\begin{figure}[ptb]
\begin{centering}
\begin{tabular}{ll}
a) & b)\tabularnewline
\includegraphics[width=8cm]{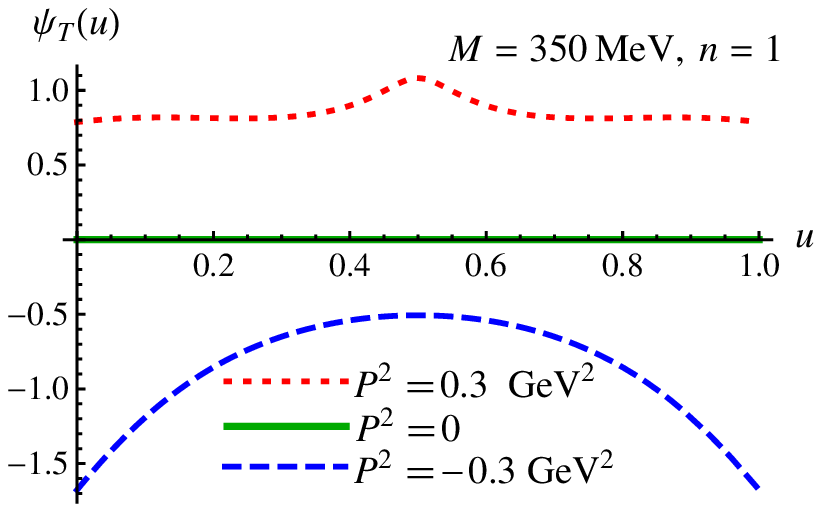} & \includegraphics[width=8cm]{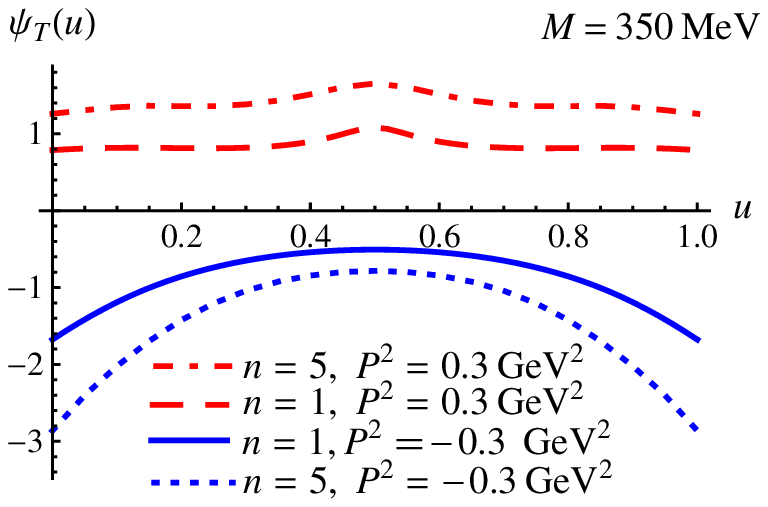}\tabularnewline
c) & d)\tabularnewline
\includegraphics[width=8cm]{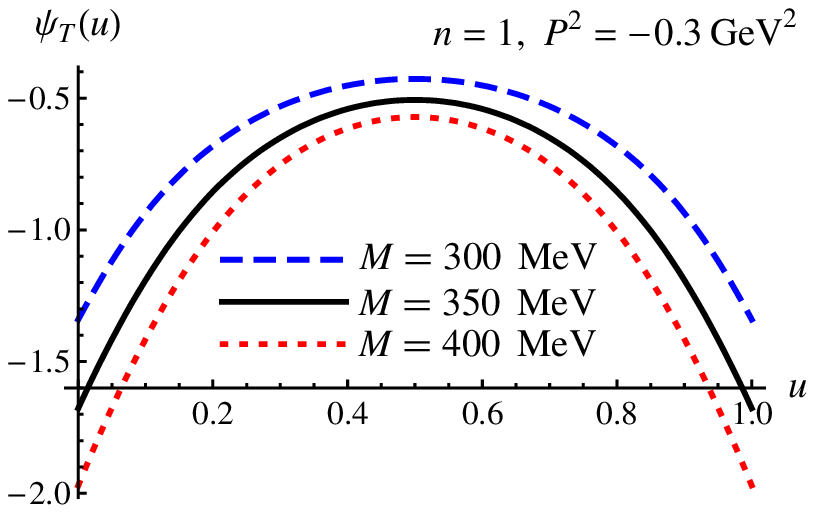} & \includegraphics[width=8cm]{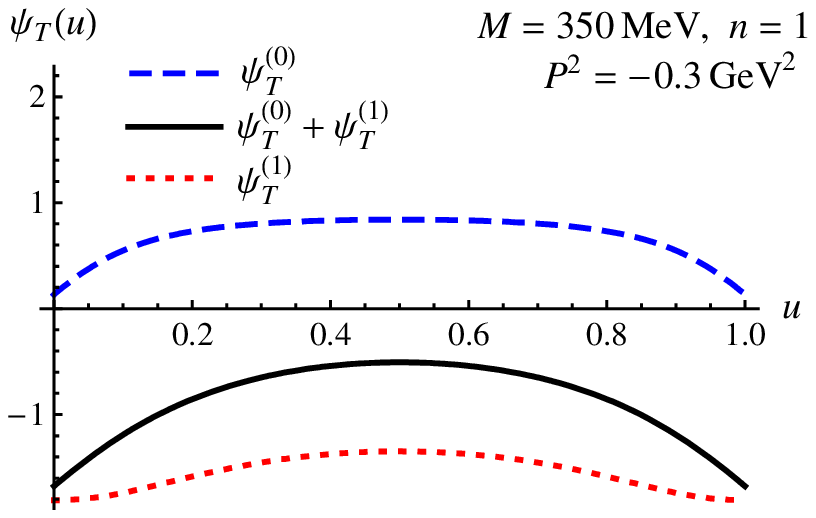}\tabularnewline
\end{tabular}
\end{centering}
\caption{Tensor twist 3 photon DA for:
a) different photon virtualities and fixed $M=350$~MeV and $n=1$
(for $P^2=0$ it is identically zero),
b) different $n$ and $P^2$ for fixed $M=350$~MeV,
c) different $M$ and fixed $n=1$ and $P^2=-0.3$~GeV$^2$,
d) decomposition into contributions corresponding to local (dashed) and non-local (dotted) parts of the vector vertex.}%
\label{fig:tensor_t3}%
\end{figure}

\begin{figure}[ptb]
\begin{centering}
\begin{tabular}{ll}
a) & b)\tabularnewline
\includegraphics[width=8cm]{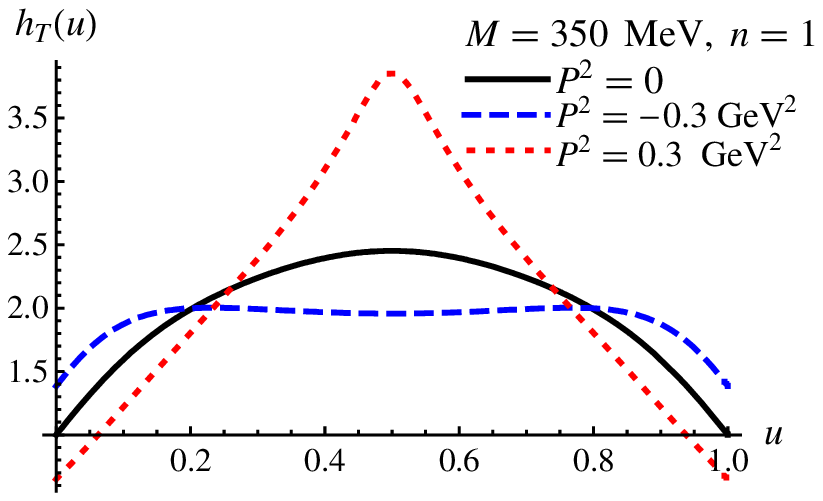} & \includegraphics[width=8cm]{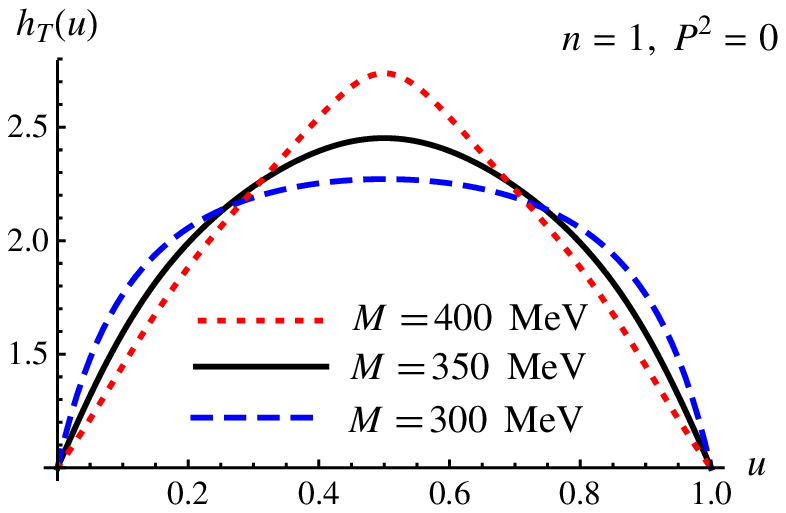}\tabularnewline
c) & d)\tabularnewline
\includegraphics[width=8cm]{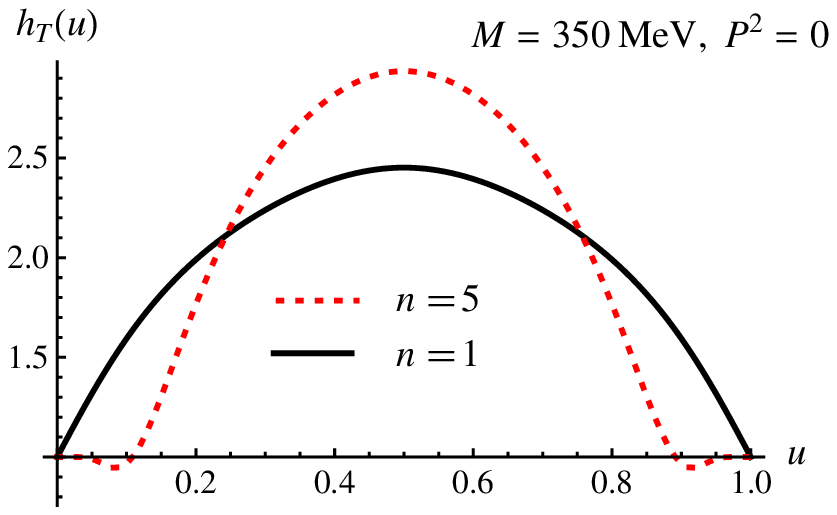} & \includegraphics[width=8cm]{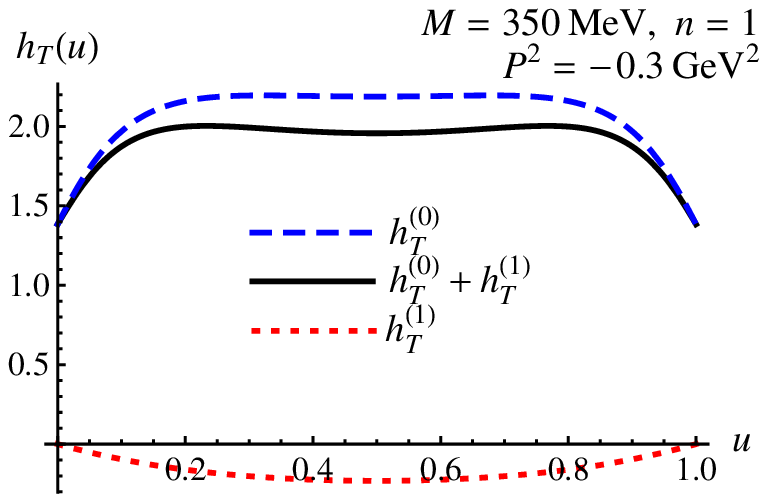}\tabularnewline
\end{tabular}
\end{centering}
\caption{Tensor twist 4 photon DA (without end point delta functions) for:
a) different photon virtualities and fixed $M=350$~MeV and $n=1$,
b) various $M$ and fixed $n=1$ and $P^2=0$,
c) two choices of $n=1,5$ and fixed $M=350$~MeV and $P^2=0$,
d) decomposition into contributions corresponding to local (dashed)
and non-local (dotted) parts of the vector vertex.}%
\label{fig:tensor_t4}%
\end{figure}

Full results (with nonlocal current) for twist 3 $\psi_{T}$ and twist 4
$h_{T}$ tensor distributions are show in figs. \ref{fig:tensor_t3} and \ref{fig:tensor_t4}
respectively.

\subsubsection{Vector DAs}

\label{vectoh}

In the case of higher twist vector DAs, $\psi_{V}$ and $h_{V}$, calculations
are basically the same, with the restriction that we have to perform
subtractions similarly to the twist 2 case. For twist 3 amplitude we obtain:
\begin{equation}
\psi_{V}(u)=\frac{i4N_{c}P^{+}}{f_{3\gamma}F_{V}(P^{2})}%
{\displaystyle\int}
\frac{d^{4}k}{(2\pi)^{4}}\frac{\left(  T_{V}^{(0)}+T_{V}^{(1)}\right)
\,\delta\left(  k\cdot n-u\,n\cdot P\right)  }{(k^{2}-M_{k}^{2})((k-P)^{2}%
-M_{k-P}^{2})}%
\end{equation}
with
\begin{align}
T_{V}^{(0)} &  =\left[  2(\vec{\varepsilon}_{\bot}\cdot\vec{k}_{\bot}%
)^{2}-\vec{k}_{\bot}^{\,2}\right]  -\varepsilon^{+}(\vec{\varepsilon}_{\bot
}\cdot\vec{k}_{\bot})\left(  k^{-}-u\frac{P^{2}}{P^{+}}\right)  \nonumber\\
&  -\frac{1}{2}(1-2u)P^{+}k^{-}-\frac{1}{2}uP^{2}-M_{k}M_{k-P}.\label{psiVtr0}%
\end{align}
In fact after integrating over the transverse angle the terms in the first
line vanish. Next%
\begin{equation}
T_{V}^{(1)}=\frac{(\vec{\varepsilon}_{\bot}\cdot\vec{k}_{\bot})}{P^{+}}%
\frac{\left(  M_{k}^{2}-M_{k-P}^{2}\right)  }{k^{-}+\bar{u}\frac{P^{2}}{P^{+}%
}}\left[  \varepsilon^{+}\left(  k^{-}-u\frac{P^{2}}{P^{+}}\right)  -2(\vec
{k}_{\bot}\cdot\vec{\varepsilon}_{\bot})\right].  \label{psiVtr1}%
\end{equation}
Again only quadratic term $(\vec{\varepsilon}_{\bot}\cdot\vec{k}_{\bot})^{2}$
survives integration over the transverse angle. The result can be split into a
regular part coming from the local part of the vertex:%
\begin{align}
\psi_{V}^{(0,\, \text{reg})}(u)   & =\frac{N_{c}}{8\pi^{2}f_{3\gamma}F_{V}(P^{2})}%
{\displaystyle\sum\limits_{i,j=1}^{4n+1}}
f_{i}\,f_{j}\,\eta_{i}^{2n}\eta_{j}^{2n}\ln\left(  \frac{1+u\bar{u}p^{2}%
-\bar{u}\eta_{i}+u\eta_{j}}{u\bar{u}p^{2}}\right)  \nonumber\\
& \left(  \left(  1+2u\bar{u}\right)  \eta_{i}^{2n}\eta_{j}^{2n}P^{2}%
+2M^{2}+\frac{{}}{{}}(1-2u)(\eta_{i}-\eta_{j})\eta_{i}^{2n}\eta_{j}%
^{2n}\Lambda^{2}\right)  ,
\end{align}
the part with delta functions also coming from the local part of the vertex:%
\begin{equation}
\psi_{V}^{(0,\,\text{delta})}(u) =\frac{-N_{c}}{8\pi^{2}f_{3\gamma}F_{V}(P^{2}%
)}\Lambda^{2}%
{\displaystyle\sum\limits_{i=1}^{4n+1}}
f_{i}\,\eta_{i}^{4n}(1+\eta_{i})\ln\left(  1+\eta_{i}\right)  \left[
\delta(u-1)\frac{{}}{{}}+\delta(u)\right]
\end{equation}
and the nonlocal part:%
\begin{align}
\psi_{V}^{(1)}(u)  & =\frac{N_{c}}{8\pi^{2}f_{3\gamma}F_{V}(P^{2})}2M^{2}%
\,\sum_{i,j=1}^{4n+1}f_{i}f_{j}(\eta_{j}^{2n}\,+\eta_{i}^{2n})\left(
1-\eta_{i}\bar{u}+\eta_{j}u+u\bar{u}r^{2}\right)  \nonumber\\
& \qquad\qquad\qquad\frac{\eta_{i}^{2n}-\eta_{j}^{2n}}{\eta_{i}-\eta_{j}}%
\;\ln\left(  1+u\bar{u}p^{2}-\bar{u}\eta_{i}+u\eta_{j}\right)
\end{align}
The part with delta functions can be rewritten as
\begin{equation}
\psi_{V}^{(0,\,\text{delta})}(u) =\frac{-1}{2 F_{V}(P^{2}%
)}\left[
\delta(u-1) + \delta(u)\right],
\end{equation}
where we used (\ref{f3g2}) and the identity
$\Lambda^2 \eta_i^{4n}(1+\eta_i)=M^2$ following from
equation (\ref{eq:Gz}) for zeros of $G(z)$.

Next we calculate twist 4 vector distribution amplitude $h_{V}(u)$:
\begin{equation}
h_{V}(u)=\frac{i4N_{c}P^{+}}{f_{3\gamma}F_{V}(P^{2})}\,\frac{\left(P^+\right)}{2P^2}%
{\displaystyle\int}
\frac{d^{4}k}{(2\pi)^{4}}\frac{\left(  R_{V}^{(0)}+R_{V}^{(1)}\right)
\,\delta\left(  k\cdot n-u\,n\cdot P\right)  }{(k^{2}-M_{k}^{2})((k-P)^{2}%
-M_{k-P}^{2})}%
\end{equation}
where the traces read%
\begin{align}
R_{V}^{(0)}  & =\left(k^-\right)^2 -\frac{P^2}{P^+}k^- -\frac{P^2}{\left(P^+\right)^2} k_T^2-M_k M_{k-P}\\
R_{V}^{(1)}  & =\frac{k^- -\frac{P^2}{P^+}u}{k^{-}+\bar{u}\frac{P^2}{P^+}}\,
\left[ k^- \left(M^2_k-M^2_{k-P}\right) - M_k\left(M_k - M_{k-P}\right)\right]
\end{align}
The only additional complication is due to the second derivative of delta
function -- the details can be found appendix \ref{light} eqs.(\ref{J20end}%
)--(\ref{J22}). Results for the local part read:%
\begin{align}
h_{V}^{(0,\,\text{reg})}(u)  & =\frac{-N_{c}}{8\pi^{2}f_{3\gamma}F_{V}(P^{2})}%
\, 2{\displaystyle\sum\limits_{i,j=1}^{4n+1}}
f_{i}\,f_{j}\,\eta_{i}^{2n}\eta_{j}^{2n}\ln\left(  \frac{1+u\bar{u}p^{2}%
-\bar{u}\eta_{i}+u\eta_{j}}{u\bar{u}p^{2}}\right)  \nonumber\\
& \bigg(  2u\bar{u}\,  \eta_{i}^{2n}\eta_{j}^{2n}P^{2}%
-2M^{2}+\eta_{i}^{2n}\eta_{j}^{2n}\left( 2\eta_i-\eta_j+1-3u(\eta_i-\eta_j)\right)
\Lambda^{2}\\ & + \frac{(\eta_i-\eta_j)^2 \Lambda^4}{P^2}\bigg)  ,
\end{align}
and%
\begin{align}
h_{V}^{(0,\,\text{delta})}(u)  & =\frac{-N_{c}}{8\pi^{2}f_{3\gamma}F_{V}(P^{2})}%
\, \frac{2\, \Lambda^2}{P^2}\bigg\{-2 M^2 {\displaystyle\sum\limits_{i=1}^{4n+1}}
f_{i}\, \ln\left( 1+\eta_i \right) \nonumber\\
& \qquad\qquad\qquad\qquad\left[(1+\eta_i)\,\left[ \delta(u) +\delta(u-1)\right]
+p^2 \delta(u)\right] \nonumber\\
& + \Lambda^2\, {\displaystyle\sum\limits_{i,j=1}^{4n+1}}
f_{i}\, f_j \eta_i^{4n}\eta_j^{4n}\frac{1}{2}\ln\left(  \frac{1+u\bar{u}p^{2}%
-\bar{u}\eta_{i}+u\eta_{j}}{u\bar{u}p^{2}}\right)
\left(1+u\bar{u}p^{2}-\bar{u}\eta_{i}+u\eta_{j}\right)^2 \nonumber\\
& \qquad\qquad\qquad\qquad\left[\delta ' (u)+\delta ' (u-1)\right]\bigg\},
\end{align}
and for the contribution coming from the nonlocal part of the photon vertex:%
\begin{align}
h_{V}^{(1,\,\text{reg})}(u)  & =\frac{-N_{c}}{8\pi^{2}f_{3\gamma}F_{V}(P^{2})}
\, \frac{2M^2\, \Lambda^2}{P^2}{\displaystyle\sum\limits_{i,j=1}^{4n+1}}
f_{i}\,f_{j}\,\ln\left(1+u\bar{u}p^{2}-\bar{u}\eta_{i}+u\eta_{j}\right) \nonumber\\
& \qquad\qquad\qquad\frac{\eta_{i}^{2n}-\eta_{j}^{2n}}{\eta_{i}-\eta_{j}}\,
\left( \eta_i - \eta_j -(2u-1)p^2\right)
\,\left( \left( \eta_i^{2n}+\eta_j^{2n}\right) - p^2\,\eta_j^{2n}\right),
\end{align}
\begin{align}
h_{V}^{(1,\,\text{delta})}(u)  & =\frac{-N_{c}}{8\pi^{2}f_{3\gamma}F_{V}(P^{2})}%
\, \frac{2M^2\, \Lambda^2}{P^2}{\displaystyle\sum\limits_{i=1}^{4n+1}}
f_{i}\,(1+\eta_{i})\ln\left( 1+\eta_i \right)
\,\left[\delta (u)+\delta (u-1)\right].
\end{align}

We note that
\begin{equation}
\int_{0}^{1} \, h_{V}^{(0,\,\text{delta})}(u)\, du
= \frac{-N_{c}}{8\pi^{2}f_{3\gamma}F_{V}(P^{2})}%
\, \frac{-4\, M^2\,\Lambda^2}{P^2}
{\displaystyle\sum\limits_{i=1}^{4n+1}}
f_{i}\,(1+\eta_{i})\ln\left( 1+\eta_i \right),
\end{equation}
thus it cancels with $\int_{0}^{1} \,  h_{V}^{(1,\,\text{ delta})}(u)\, du$ as can be easily seen.

\begin{figure}[ptb]
\begin{centering}
\begin{tabular}{ll}
a) & b)\tabularnewline
\includegraphics[width=8cm]{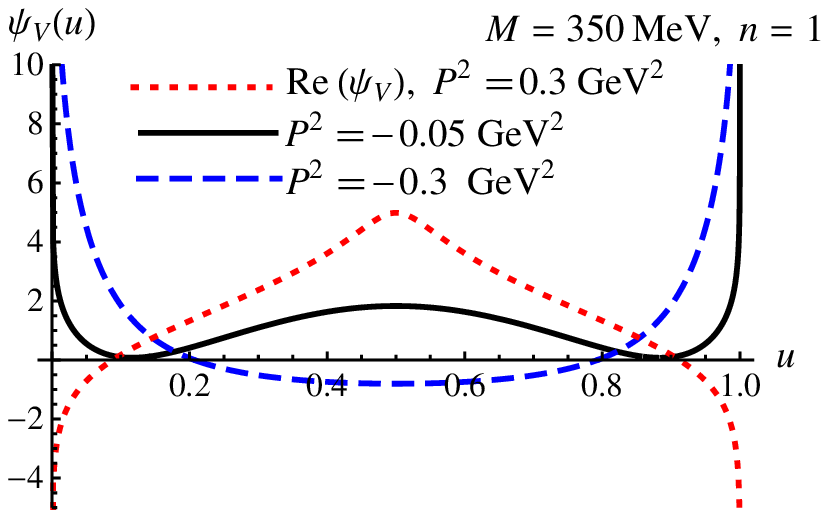} & \includegraphics[width=8cm]{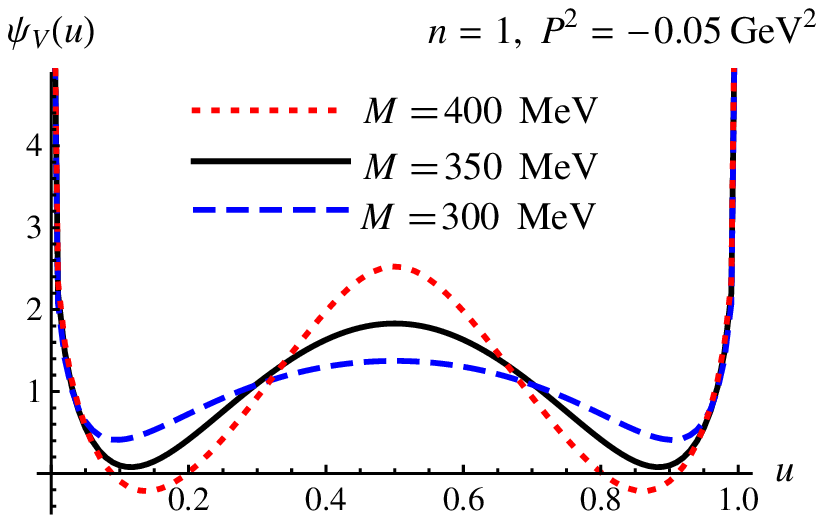}\tabularnewline
c) & d)\tabularnewline
\includegraphics[width=8cm]{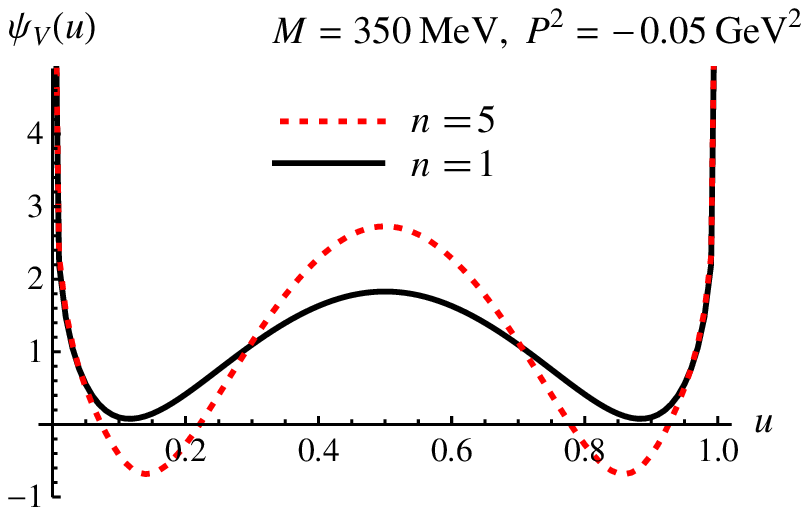} & \includegraphics[width=8cm]{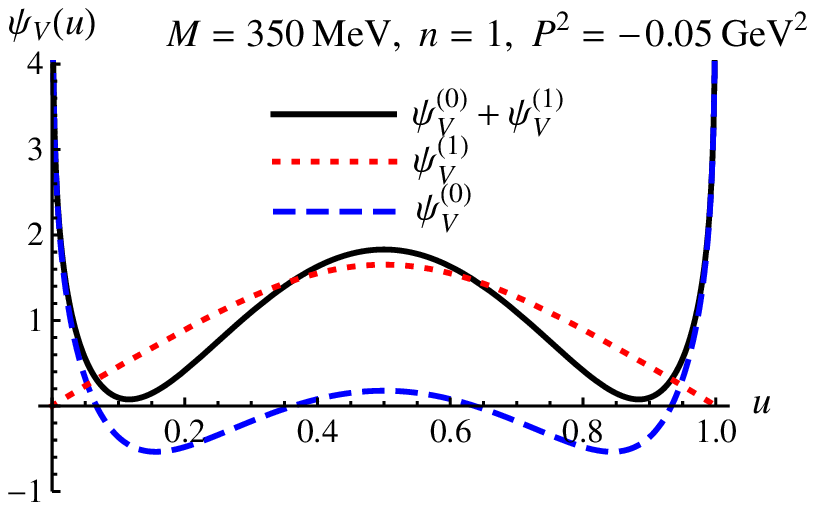}\tabularnewline
\end{tabular}
\end{centering}
\caption{Vector twist 3 photon DA for:
a) different photon virtualities and fixed $M=350$~MeV and $n=1$,
b) various $M$ and fixed $n=1$ and $P^2=-0.05$~GeV$^2$,
c) two choices of $n=1,5$ and fixed $M=350$~MeV and
$P^2=-0.05$~GeV$^2$,
d) decomposition into contributions corresponding to local (dashed)
and non-local (dotted) parts of the vector vertex.}%
\label{fig:vector_t3}%
\end{figure}

\begin{figure}[ptb]
\begin{centering}
\begin{tabular}{ll}
a) & b)\tabularnewline
\includegraphics[width=8cm]{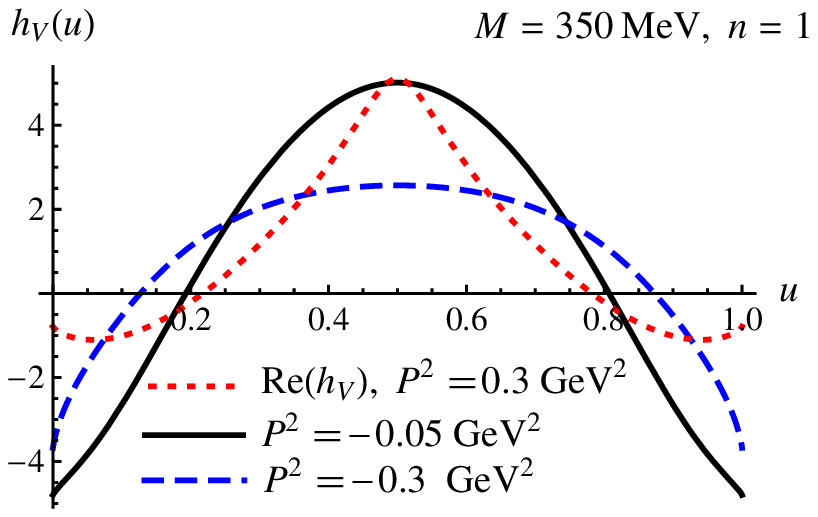} & \includegraphics[width=8cm]{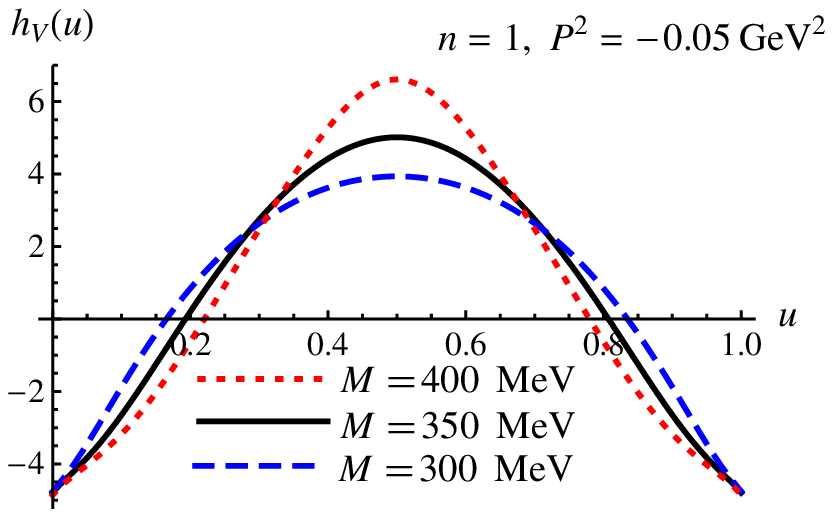}\tabularnewline
c) & d)\tabularnewline
\includegraphics[width=8cm]{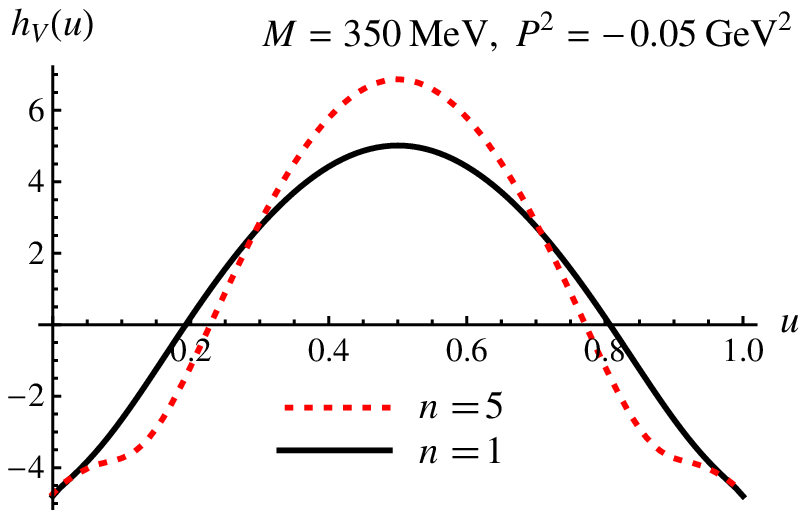} & \includegraphics[width=8cm]{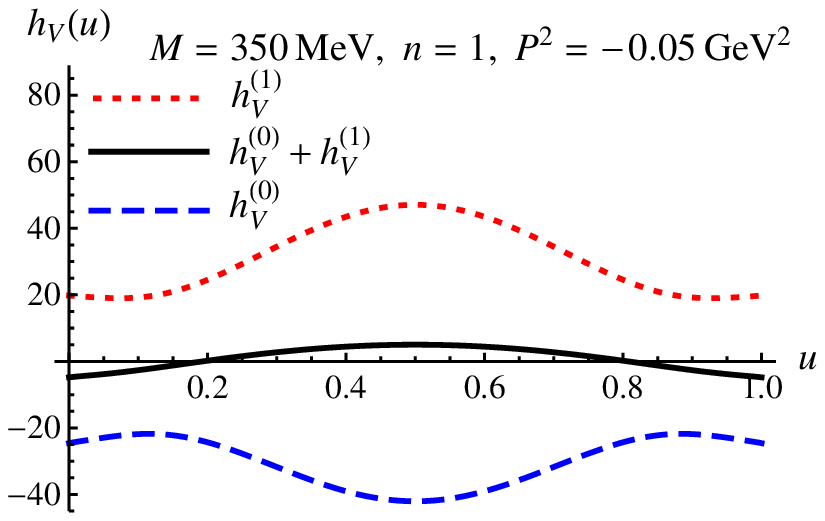}\tabularnewline
\end{tabular}
\end{centering}
\caption{Vector twist 4 photon DA for:
a) different photon virtualities and fixed $M=350$~MeV and $n=1$,
b) various $M$ and fixed $n=1$ and $P^2=-0.05$~GeV$^2$,,
c) two choices of $n=1,5$ and fixed $M=350$~MeV
and $P^2=-0.05$~GeV$^2$,
d) decomposition into contributions corresponding to local (dashed)
and non-local (dotted) parts of the vector vertex.}%
\label{fig:vector_t4}%
\end{figure}

Our results are shown in figs. \ref{fig:vector_t3} and \ref{fig:vector_t4} for twist 3 and
twist 4 respectively. The magnitude of
both distributions is growing unlimitedly when photon becomes softer (obviously distributions multiplied
by the vector form factor remain finite). Notice
however that for the real photon $h_{V}$ decouples.

\subsubsection{Axial DA}

\label{axialh}

We have only one distribution in the axial vector channel which is of twist 3.
When inverting the definition \eqref{eq:def_axial}, due to presence of
$\lambda$ on the right hand side, we obtain the expression for the derivative
of DA rather then for DA itself%
\begin{align}
\tilde{\psi}_{A}^{\prime}\left(  u\right)   &  =\psi_{A}^{\prime}\left(
u\right)  +\psi_{A}\left(  0\right)  \delta\left(  u\right)  -\psi_{A}\left(
1\right)  \delta\left(  \bar{u}\right)  \label{eq:axial1}\\
&  =-i\frac{8N_{c}}{f_{3\gamma}\,F_{A}(P^{2})}\int\frac{d^{D}k}{\left(
2\pi\right)  ^{D}}\frac{T_{A}^{(0)}}{(k^{2}-M_{k}^{2})((k-P)^{2}-M_{k-P}^{2}%
)}\delta\left(  k\cdot n-uP^{+}\right)  ,\nonumber
\end{align}
with%
\begin{equation}
T_{A}^{(0)}=-\frac{P^{+2}}{2}\left(  k^{-}-\frac{P^{2}}{P^{+}}u\right)
-\varepsilon^{+}P^{2}\,(\vec{\varepsilon}_{\bot}\cdot\vec{k}_{\bot}).
\end{equation}
In the case of axial DA the nonlocal part of the vertex does not give any
contribution. This is simply because the Dirac trace is equal to zero. To
obtain $\psi_{A}\left(  u\right)  $ one has to integrate \eqref{eq:axial1}
over $du^{\prime}$ from $0$ to $u$:%
\begin{equation}
\psi_{A}\left(  u\right)  =%
{\displaystyle\int\limits_{0}^{u}}
\tilde{\psi}_{A}^{\prime}\left(  u^{\prime}\right)  du^{\prime}%
.\label{eq:axial2}%
\end{equation}
Notice that the end point contributions cancel out and one might get an
impression that $\psi_{A}\left(  u\right)  $ is determined up to a constant.
Fortunately we have at our disposal an independent formula for $F_{A}(P^{2})$
given by eq. (\ref{eq:ax_ff}) (see also eq. (\ref{eq:aff9}) in appendix
\ref{axialf}) and the normalization condition (\ref{eq:normaliz3}) that fix
the value of $\psi_{A}(0)=\psi_{A}(1)\neq0$ at nonzero value.

As in the case of vector twist 2 DA $\psi_{A}(u)$ is UV divergent and requires
subtraction of the perturbative part. The result splits into a regular part%
\begin{align}
\,\tilde{\psi}_{A}^{(a)\,\prime}(u)  &  =\frac{N_{c}}{4\pi^{2}f_{3\gamma
}\,F_{A}(P^{2})}%
{\displaystyle\sum\limits_{i,j=1}^{4n+1}}
f_{i}\,f_{j}\,\eta_{i}^{4n}\eta_{j}^{4n}\left(  \left(  \eta_{i}-\eta
_{j}\right)  \Lambda^{2}+(1-2u)P^{2}\right) \\
&  \,\,\,\qquad\qquad\,\,\,\qquad\qquad\,\,\,\qquad\ln\left(  \frac{1+u\bar
{u}p^{2}-\bar{u}\eta_{i}+u\eta_{j}}{u\bar{u}p^{2}}\right) \nonumber
\end{align}
and the piece involving $\delta$ functions:%
\begin{equation}
\,\tilde{\psi}_{A}^{(a)\,\prime}(u)=\frac{N_{c}}{4\pi^{2}f_{3\gamma}%
\,F_{A}(P^{2})}\Lambda^{2}%
{\displaystyle\sum\limits_{i=1}^{4n+1}}
f_{i}\,\eta_{i}^{4n}\,(1+\eta_{i})\ln\left(  1+\eta_{i}\right)  \left[
\delta(u)\frac{{}}{{}}-\delta(u-1)\right]  .
\end{equation}
Note that $\eta_{i}^{4n}\,(1+\eta_{i})=r^{2}$, and in virtue of (\ref{f3g2})%
\begin{equation}
\,\tilde{\psi}_{A}^{(a)\,\prime}(u)=\frac{1}{\,F_{A}(P^{2})}\left[
\delta(u)\frac{{}}{{}}-\delta(u-1)\right]  .
\end{equation}

\begin{figure}[ptb]
\begin{centering}
\begin{tabular}{ll}
a) & b)\tabularnewline
\includegraphics[width=8cm]{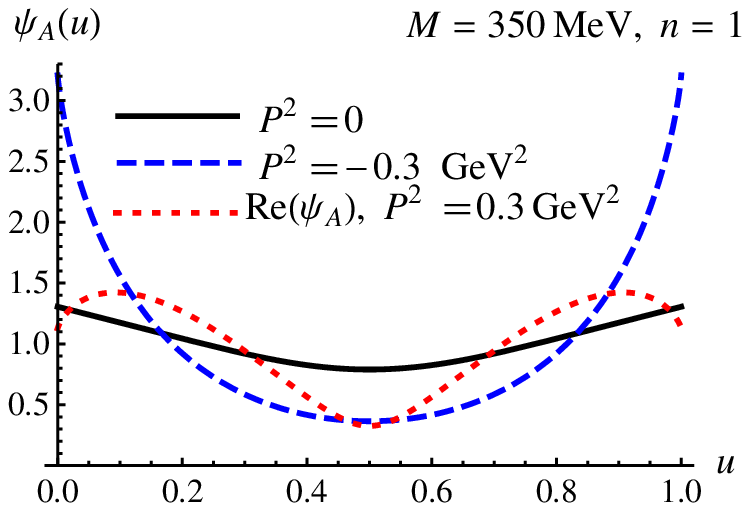} & \includegraphics[width=8cm]{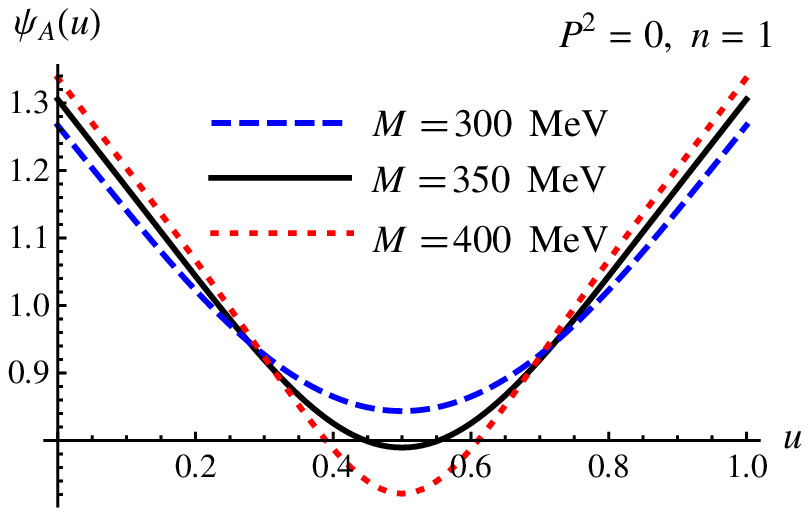}\tabularnewline
c) & d)\tabularnewline
\includegraphics[width=8cm]{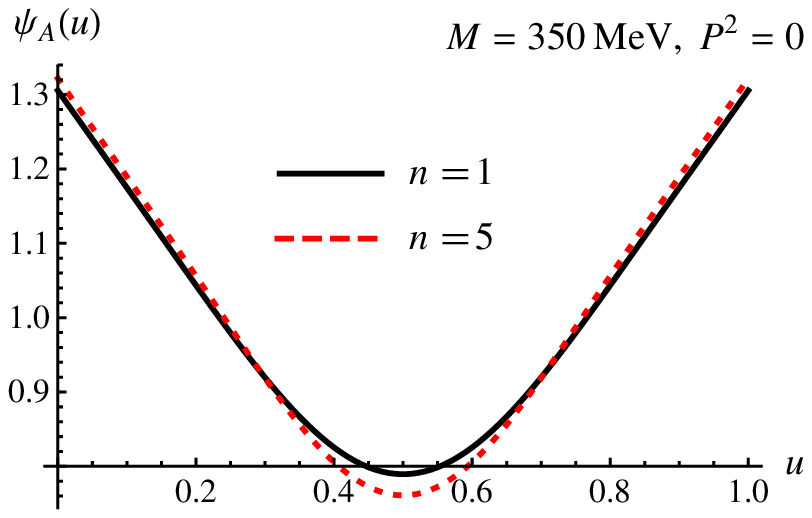} & \includegraphics[width=8cm]{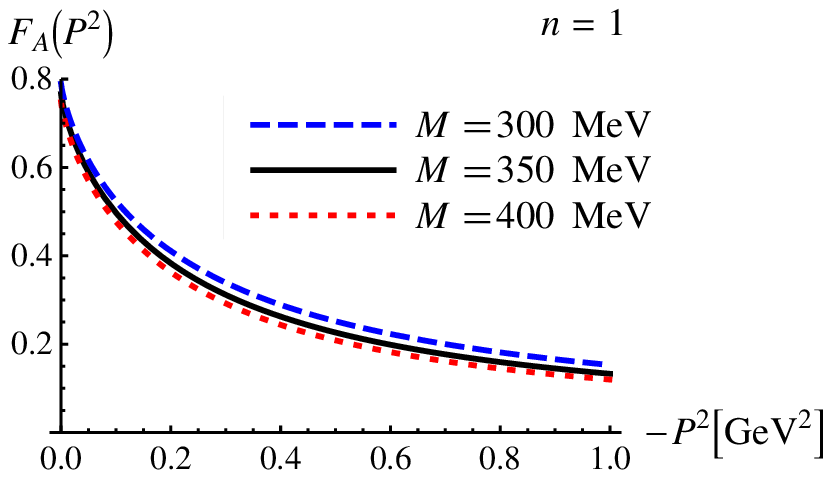}\tabularnewline
\end{tabular}
\end{centering}
\caption{Axial photon DA for:
a) different photon virtualities and fixed $M=350$~MeV and $n=1$,
b) various $M$ and fixed $n=1$ and $P^2=0$,
c) two choices of $n=1,5$ and fixed $M=350$~MeV and $P^2=0$.
d) Axial form factor for different choices of $M$ (there is almost no $n$ dependence).}%
\label{fig:axial}%
\end{figure}

The form factor $F_{A}(P^{2})$ and $\psi_{A}(u)$ itself is shown in fig. \ref{fig:axial}. We obtain the following
values for axial form factor at zero momenta: $F_{A}\left(  0\right)
\approx0.77$ for $M=350\,\mathrm{MeV}$, $F_{A}\left(  0\right)  \approx0.79$
for $M=300\,\mathrm{MeV}$ and $F_{A}\left(  0\right)  \approx0.75$ for
$M=400\,\mathrm{MeV}$.

\section{Summary}

\label{summa}

In this work we calculated analytically a set of photon distribution
amplitudes up to twist four in tensor, vector and axial vector channels. We
used nonlocal chiral quark model with momentum dependent quark mass. In order
to get a correct behavior of low energy matrix elements we modified vector
vertices (making them nonlocal) in such a way that Ward-Takahashi identities
were fulfilled (\ref{eq:vertex}). Similar, although numerical calculation was
already done in {ref. \cite{Dorokhov:2006qm}}. They also used instanton
motivated nonlocal model with dressed vertices, taking into account
rescattering in the $\rho$ meson channel. The shape of the mass dependence on
momentum was chosen as an exponent decreasing with $k^{2}$. Here we use
$F\left(  k\right)  $ as given by \eqref{Fkdef} and neglect rescattering which
turns out to be small.

First we obtained numerical estimates for quark condensate $\left\langle
\bar{\psi}\psi\right\rangle $, magnetic susceptibility $\chi_{m}$ and decay
constant $f_{3\gamma}$ in our model. For larger values of constituent quark
mass $M$ or power $n$ our results are getting close to the ones of ref.
\cite{Dorokhov:2006qm}.
Unlike $\left\langle \bar{\psi}\psi\right\rangle $ and $\chi_{m}$
the value of $f_{3\gamma}$ is rather stable as far as model parameters are
concerned. Using evolution equations (following refs. \cite{Ball:2002ps,Dorokhov:2006qm}) we find that for
$M=350\,\mathrm{MeV}$ and $n=5$, the scale of our model is about $\mu
\approx 500\,\mathrm{MeV}$ (this estimation was done using
$\left\langle \bar{\psi}\psi\right\rangle $, $\chi_{m}$, $f_{3\gamma}$ as
given by sum rules at $1\,\mathrm{GeV}$ scale and evolving them backwards
down to the model values). This is in rough agreement
with the scale of the instanton liquid model which is believed to be of the
order of 600 MeV \cite{Diakonov:1983hh}.

Next let us discuss the properties of the distribution amplitudes obtained
within the present approach. Leading twist amplitudes are not very sensitive
to the value of power $n$. However it seems that higher twist DAs are rather
strongly model dependent.

Comparing our results with those of ref. \cite{Dorokhov:2006qm} we find some
similarities, but also some discrepancies. Tensor leading twist DAs are in
fact the same. For real photons they are almost constant with small maximum at
$u=1/2$ and they do not vanish at the end points. The contribution of the
nonlocal part of the vertex is rather small, it is however producing the small
maximum in the middle. For $P^{2}\neq0$ the end points move up for negative
$P^{2}$ and down for positive $P^{2},$ whereas the middle value behaves in the
opposite way.

Twist 2 vector DA ($\phi_{V}$) should decouple for $P^{2}=0$. Here the
importance of gauge invariance shows up. We find cancelation of two
contributions to $\phi_{V}$ coming from the local and nonlocal parts of the
photon vertex which are not proportional to $P^{2}$. The remaining part is
therefore proportional to $P^{2}$ and decouples as required by the gauge
invariance. We find that $\phi_{V}$ vanishes at the end points and develops
minimum in the middle for $P^{2}\ll0$, whereas for $P^{2}\gg0$ it has a
bell-like shape with a small dip in the middle. This behavior is very different from
the one obtained in ref. \cite{Dorokhov:2006qm} where $\phi_{V}$ is almost
flat and does not vanish at the end points. However both vector and also
tensor form factors are quite similar in both cases.

One has to note that because of
the subtraction of the perturbative part that is required in this case,
$\phi_{V}$ develops imaginary part for positive photon virtualities, so in
this case we only discuss the real part.

As far as higher twist DAs are concerned the situation is as follows.

Our tensor twist 3 DA ($\psi_T$) is identically zero for $P^2=0$, because it is
simply proportional to  $P^2$. For $P^2<0$ it is negative and has the shape
of inverted "U", similarly to the one of ref. \cite{Dorokhov:2006qm} .
In this case  DA is a regular function without $\delta$-type singularities.
Vector twist 3 DA  ($\psi_V$) in our case blows up at the end points;
such behavior is not seen
in ref. \cite{Dorokhov:2006qm}. However, similarly to
ref. \cite{Dorokhov:2006qm} we also obtained delta-type singularities
at the edges of the physical support. Twist 3 axial DAs ($\psi_A$) in both
cases show similar behavior: they do not vanish at the end points and
have a minimum for $u=1/2$. Despite the fact that for $P^{2}=0$
axial vector DAs  both in our case and
in the case of ref. \cite{Dorokhov:2006qm} look similar, the axial form
factors behave differently for $P^{2}<0$. In our case $F_{A}(P^{2})$ vanishes
at large negative momenta in contrary to the one of 
ref. \cite{Dorokhov:2006qm} that tends to unity in the same limit.

Regular part (without delta-type singularities) of
twist 4 tensor DA ($h_T$) is in our case positive and vanishes at the
end points for  $P^2=0$ whereas in ref. \cite{Dorokhov:2006qm}
it is negative and does not vanish at the end points. For space-like
photon momentum $P^2<0$ we see some similarity in shape
between our $h_T$ and $-h_T$ of ref. \cite{Dorokhov:2006qm}.
Vector twist 4 DA ($h_V$) is in our case a result of large cancelation
of the positive non-local piece and the negative local piece.
Its properties are not discussed in detail in ref. \cite{Dorokhov:2006qm}.

The only phenomenologically accessible photon distribution amplitude
is the leading twist tensor DA -- $\phi_T$. It is almost flat and does not
vanish at the end points. This behavior is seen in our model and in other
models discussed in ref. \cite{Dorokhov:2006qm} and also in
refs. \cite{Praszalowicz:2001wy,Petrov:1998kg}. Flat DA is
characteristic for the elementary point-like particle, however, it is
violating factorization theorems of QCD that require the DAs to
vanish for $u=0,1$. Formal evolution of such an amplitude is questionable,
not only because the Gegenbauer series is not convergent at the end points,
but also, because potentially large contributions coming from the
vicinity of $u=0,1$ are not summed by the ERBL evolution equations
\cite{Radyushkin}.

\begin{acknowledgments}
The authors are grateful to I. Anikin, W. Broniowski, A. Dorokhov
and E. Ruiz-Arriola for
discussions. The paper was partially supported by the Polish-German
cooperation agreement between Polish Academy of Science and DFG.
\end{acknowledgments}

\appendix

\section{Identities}

\label{ident}

In this appendix we summarize some of the identities used in this paper that
deal with the sums of factors $f_{i}$ and powers of $\eta_{i}$. Some of them
have been already introduced in ref. \cite{Praszalowicz:2001wy}, but the
general proofs have not been given. For definiteness let us recall that
$\eta_{i},$ $i=1,\ldots,4n+1$ are the solutions of the algebraic equation
$G\left(  z\right)  =z^{4n+1}+z^{4n}-r^{2}=0$. We denote
\[
f_{i}=\prod_{j \ne i}\left(  \eta_{i}-\eta_{j}\right)  ^{-1}.
\]
\qquad

For \textit{any} set of $4n+1$ complex numbers $\eta_{i}$ (not necessarily
satisfying $G\left(  z\right)  =0$) we have%
\begin{equation}
\sum_{i=1}^{4n+1}f_{i}\eta_{i}^{N}=%
\begin{cases}
1 & \mathrm{for}\,N=4n,\\
0 & \mathrm{for\,}N<4n.
\end{cases}
\label{eq:th1}%
\end{equation}
To prove (\ref{eq:th1}) let's define the following function%
\begin{equation}
f\left(  z\right)  =\frac{z^{M}}{\prod_{i=1}^{4n+1}\left(  z-\eta_{i}\right)
}, \label{eq:th1a}%
\end{equation}
where $M\leq4n$ and integrate it over a circle with infinite radius. On one
hand we can use residue technique to get the sum entering (\ref{eq:th1}),
on the other hand, direct integration over the large circle gives right hand side
of (\ref{eq:th1}).

If, in addition, $\eta_{i}$ satisfies $G\left(  \eta_{i}\right)  =0$ then%
\begin{equation}
\sum_{i=1}^{4n+1}f_{i}\eta_{i}^{P}=\left(  -1\right)  ^{P} \label{eq:th2}%
\end{equation}
for $4n\leq P\leq8n$. This can be proven in the following way. Notice that for
$P=4n$ equality (\ref{eq:th2}) is satisfied due to \eqref{eq:th1}. Let us move
to $P=4n+1$, that is we want to calculate%
\begin{equation}
x=\sum_{i=1}^{4n+1}f_{i}\eta_{i}^{4n+1}. \label{eq:th2a}%
\end{equation}
Adding to this equation the result for $P=4n$ we get
\begin{equation}
\sum_{i=1}^{4n+1}f_{i}\left(  \eta_{i}^{4n+1}+\eta_{i}^{4n}\right)  =x+1.
\label{eq:th2b}%
\end{equation}
Using the fact that $G\left(  \eta_{i}\right)  =\eta_{i}^{4n+1}+\eta_{i}%
^{4n}-r^{2}=0$ and \eqref{eq:th1} we have%
\begin{equation}
x+1=r^{2}\sum_{i=1}^{4n+1}f_{i}=0. \label{eq:th2c}%
\end{equation}
and $x=-1$. This procedure can be repeated several times to prove
(\ref{eq:th2}) until $P=8n$.

For any set of $4n+1$ complex numbers $\eta_{i}$ we have the following
identity%
\begin{equation}
\sum_{i=1}^{4n+1}f_{i}\eta_{i}^{4n+1}=\sum_{i=1}^{4n+1}\eta_{i}.
\label{eq:th3}%
\end{equation}
Imposing in addition the constraint $G\left(  \eta_{i}\right)  =0$ we get
$\sum_{i}^{4n+1}\eta_{i}=-1.$ The proof is similar to the one of
\eqref{eq:th1}, however we integrate the following function%
\begin{equation}
g\left(  z\right)  =\frac{z^{4n+1}}{\left(  z-\eta_{1}\right)  \left(
z-\eta_{2}\right)  \ldots\left(  z-\eta_{i}\right)  ^{2}\ldots\left(
z-\eta_{4n+1}\right)  }. \label{eq:th3a}%
\end{equation}
After several algebraic steps we arrive at (\ref{eq:th3}).

\section{Pion decay constant}

\label{piond}

In our model Birse-Bowler formula \eqref{eq:Fpi} for pion decay constant
reduces to the following form%
\begin{multline}
F_{\pi}^{2}=-\frac{N_{c}M^{2}}{4\pi^{2}}\sum_{i,j=1}^{4n+1}f_{i}f_{j}\eta
_{i}^{2n} \big(\left(  1+2n\left(  1+2n\right)  \right)  \eta_{j}%
^{2n+1}+\left(  1+4n\left(  1+3n\right)  \right)  \eta_{j}^{2n}\\
+2n\left(  1+6n\right)  \eta_{j}^{2n-1}+4n^{2}\eta_{j}^{2n-2}\big)\bigg(\frac
{\epsilon_{ij}}{\eta_{i}-\eta_{j}} \left(  \ln\left(  1+\eta_{i}\right)
-\ln\left(  1+\eta_{j}\right)  \right)  +\frac{\delta_{ij}}{1+\eta_{i}%
}\bigg) \label{eq:Fpi1}%
\end{multline}
where $\epsilon_{ij}$ is $0$ for $i=j$ and $1$ otherwise, while $\delta_{ij}$
is Kronecker delta.

\section{Light-cone integrals in Schwinger representation}

\label{light}

Here we  summarize formulae used to perform $d^{D}k$ loop integration
in the presence of $\delta(n\cdot\kappa-u\,n\cdot p)$. We will consider three
cases when the numerator contains no $k^{-}$ at all and one or two powers of
$k^{-}$. We follow closely the method of ref. \cite{BroniowskiTDA}.

Consider loop integral (\ref{loop2}) and apply to it (\ref{decomposition}):%
\begin{equation}
\mathcal{J}=\mathcal{A}\Lambda^{D-5}\int\frac{d^{D}\kappa}{\left(
2\pi\right)  ^{D}}\delta(n\cdot\kappa-u\,n\cdot p)%
{\displaystyle\sum\limits_{i,j=1}^{4n+1}}
f_{i}f_{j}\frac{\eta_{i}^{4n}\eta_{j}^{4n}\mathcal{N}}{(z_{1}-\eta_{i}%
)(z_{2}-\eta_{j})}%
\end{equation}
expressed in terms of scaled variables (\ref{eq:dimensionless_var}). Here
$\mathcal{N}$ is the numerator to be specified later. Recall that
\begin{equation}
z_{1}=\left(  \kappa-p\right)  ^{2}-1+i\epsilon,\;z_{2}=\kappa^{2}%
-1+i\epsilon.
\end{equation}
We shall now make continuation to the Euclidean metric:%
\begin{equation}
\kappa^{0}=i\kappa^{4}%
\end{equation}
with%
\begin{equation}
\kappa^{2}\rightarrow-\vec{\kappa}^{2},\;\;\kappa\cdot p\rightarrow
-\vec{\kappa}\cdot\vec{p},\;\;n\cdot\kappa\rightarrow-\vec{n}\cdot\vec{\kappa}%
\end{equation}
where arrows denote $D$ dimensional Euclidean vectors. Therefore%
\begin{equation}
\mathcal{J}=i\mathcal{A}\Lambda^{D-5}\int\frac{d^{D}\vec{\kappa}}{\left(
2\pi\right)  ^{D}}\delta(\vec{n}\cdot\vec{\kappa}+up^{+})%
{\displaystyle\sum\limits_{i,j=1}^{4n+1}}
f_{i}f_{j}\frac{\eta_{i}^{4n}\eta_{j}^{4n}\mathcal{N}}{(\vec{\kappa}%
^{2}+1+\eta_{i})((\vec{\kappa}-\vec{p})^{2}+1+\eta_{j})}.
\end{equation}
We shall parametrize now%
\begin{equation}
\frac{1}{(\vec{\kappa}^{2}+1+\eta_{i})((\vec{\kappa}-\vec{p})^{2}+1+\eta_{j}%
)}=%
{\displaystyle\int\limits_{0}^{\infty}}
d\alpha%
{\displaystyle\int\limits_{0}^{\infty}}
d\beta\,e^{-\alpha(\vec{\kappa}^{2}+1+\eta_{i})-\beta((\vec{\kappa}-\vec
{p})^{2}+1+\eta_{j})}%
\end{equation}
and%
\begin{equation}
\delta(\vec{n}\cdot\vec{\kappa}+up^{+})=%
{\displaystyle\int\limits_{-\infty}^{\infty}}
\frac{d\lambda}{2\pi}e^{-i\lambda(\vec{n}\cdot\vec{\kappa}+up^{+})}.
\end{equation}
It is convenient to introduce new variables:%
\begin{equation}
\alpha+\beta=s,\;\beta=ys,\;\alpha=(1-y)s=-\bar{y}s.
\end{equation}
Integration measure reads then%
\begin{equation}%
{\displaystyle\int\limits_{0}^{\infty}}
d\alpha%
{\displaystyle\int\limits_{0}^{\infty}}
d\beta=%
{\displaystyle\int\limits_{0}^{\infty}}
s\,ds%
{\displaystyle\int\limits_{0}^{1}}
dy.
\end{equation}
Finally we will shift momentum$\quad$%
\begin{equation}
\vec{\kappa}=\vec{\kappa}^{\,\prime}+\left(  y\,\vec{p}-i\frac{\lambda}%
{2s}\,\vec{n}\right)  .
\end{equation}
In these new variables we have%
\begin{align}
\mathcal{J}  &  =i\mathcal{A}\Lambda^{D-5}%
{\displaystyle\sum\limits_{i,j=1}^{4n+1}}
f_{i}\,f_{j}\,\eta_{i}^{4n}\eta_{j}^{4n}%
{\displaystyle\int\limits_{0}^{1}}
dy\int\frac{d\lambda}{2\pi}e^{-i\lambda p^{+}(u-y)}\nonumber\\
&
{\displaystyle\int\limits_{0}^{\infty}}
s\,dse^{-s\left[  1-\bar{y}\eta_{i}+y\eta_{j}+y\bar{y}\,p^{2}\right]  }%
\int\frac{d^{D}\vec{\kappa}^{\,\prime}}{\left(  2\pi\right)  ^{D}}%
\mathcal{N}e^{-s\vec{\kappa}^{\,\prime\,2}}. \label{final0}%
\end{align}

Further calculations depend on the nature of $\mathcal{N}$. If $\mathcal{N}$
can be expressed entirely in terms of $z_{1,2}$ then, in virtue of
(\ref{decomposition}), it is enough to replace pertinent powers of
$z_{1,2}^{N}\rightarrow\eta_{i,j}^{N}$ and perform Gaussian integration over
$\kappa^{\prime}$. Let us denote such an integral as $\mathcal{J}_{0}$. Also
integral over $d\lambda$ is trivial. In the following we shall need also
integrals with $\lambda$ and $\lambda^{2}$ which read:%
\begin{equation}
\int\frac{d\lambda}{2\pi}\{1,\lambda,\lambda^{2}\}e^{-i\lambda p^{+}%
(u-y)}=\frac{1}{p^{+}}\left\{  1,-\frac{i}{p^{+}}\partial_{y},-\frac
{1}{p^{+\,2}}\partial_{y}^{2}\right\}  \delta(u-y). \label{indel}%
\end{equation}
Hence (for $D=4-2\varepsilon$) we get%
\begin{equation}
\mathcal{J}_{0}=i\mathcal{A}\left(  \frac{1}{4\pi}\right)  ^{2-\varepsilon
}\frac{\Lambda^{-1-2\varepsilon}}{p^{+}}%
{\displaystyle\sum\limits_{i,j=1}^{4n+1}}
f_{i}\eta_{i}^{4n}\,f_{j}\eta_{j}^{4n}\mathcal{N}(\eta_{i},\eta_{j})%
{\displaystyle\int\limits_{0}^{\infty}}
\,dss^{\varepsilon-1}e^{-s\left[  1-\bar{u}\eta_{i}+u\eta_{j}+u\bar{u}%
\,p^{2}\right]  }.
\end{equation}
In order to perform the integral over $ds$ we shall use%
\begin{equation}%
{\displaystyle\int\limits_{0}^{\infty}}
\,dss^{\varepsilon-1-n}e^{-s\left[  \cdots\right]  }=\left[  \cdots\right]
^{n-\varepsilon}\Gamma(\varepsilon-n)\simeq\frac{\left[  \cdots\right]
^{n-\varepsilon}}{\varepsilon\left(  \varepsilon-1\right)  \ldots
(\varepsilon-n)}e^{-\gamma\varepsilon}+\ldots
\end{equation}
arriving at%
\begin{align}
\mathcal{J}_{0}  &  =i\frac{\mathcal{A}}{16\pi^{2}P^{+}}\left(  \frac{4\pi
e^{-\gamma}}{\Lambda^{2}}\right)  ^{\varepsilon}\frac{1}{\varepsilon}%
{\displaystyle\sum\limits_{i,j=1}^{4n+1}}
f_{i}\eta_{i}^{4n}\,f_{j}\eta_{j}^{4n}\frac{\mathcal{N}}{\left[  1-\bar{u}%
\eta_{i}+u\eta_{j}+u\bar{u}\,p^{2}\right]  ^{\varepsilon}}\nonumber\\
&  =i\frac{\mathcal{A}}{16\pi^{2}P^{+}}\left(  \frac{4\pi e^{-\gamma}}%
{\Lambda^{2}}\right)  ^{\varepsilon}%
{\displaystyle\sum\limits_{i,j=1}^{4n+1}}
f_{i}\eta_{i}^{4n}\,f_{j}\eta_{j}^{4n}\mathcal{N}\left(  \frac{1}{\varepsilon
}-\ln\left[  1-\bar{u}\eta_{i}+u\eta_{j}+u\bar{u}\,p^{2}\right]  \right)  .
\label{J0end}%
\end{align}

If numerator $\mathcal{N}$ involves additionally one power of $\kappa^{\mu}$
we have then%
\begin{equation}
\mathcal{N\rightarrow N}(\eta_{i},\eta_{j})(w\cdot\kappa)=-\mathcal{N}%
(\eta_{i},\eta_{j})(\vec{w}\cdot\vec{\kappa})
\end{equation}
where $w$ is a constant four-vector. Let us denote such an integral by
$\mathcal{J}_{1}$. Here the only difference from the previous case comes from
the integration over $\kappa^{^{\prime}}$. Since%
\begin{equation}
\vec{w}\cdot\vec{\kappa}=\vec{w}\cdot\vec{\kappa}^{\,\prime}+\left(
y\,\vec{w}\cdot\vec{p}-i\frac{\lambda}{2s}\,\vec{w}\cdot\vec{n}\right)
\label{shift1}%
\end{equation}
only the terms in parenthesis survive. After integrating over $d\lambda$ with
the help of (\ref{indel}) and over $dy$ (in the case of $\delta^{\prime}$ we
have to integrate by parts) we arrive, back in the Minkowski metric, at:%
\begin{align}
\mathcal{J}_{1}  &  =i\frac{\mathcal{A}}{16\pi^{2}P^{+}}\left(  \frac{4\pi
e^{-\gamma}}{\Lambda^{2}}\right)  ^{\varepsilon}\frac{1}{\varepsilon}%
{\displaystyle\sum\limits_{i,j=1}^{4n+1}}
f_{i}\,f_{j}\,\frac{z_{i}^{4n}z_{j}^{4n}\,\mathcal{N}}{\left[  1-\bar{u}%
z_{i}+uz_{j}+u\bar{u}\,p^{2}\right]  ^{\varepsilon}}\nonumber\label{J1end}\\
&  \qquad\left\{  u\,(w\cdot P)\,+\frac{(w\cdot n)\,}{2P^{+\,}}\left(  \left(
\eta_{i}-\eta_{j}\right)  \Lambda^{2}\frac{{}}{{}}+(1-2u)P^{2}\right)  \right.
\nonumber\\
&  \qquad\left.  -\frac{(w\cdot n)\,}{2P^{+\,}}\frac{\Lambda^{2}}%
{\varepsilon-1}\left[  (1+\eta_{j})\delta(u-1)\frac{{}}{{}}-(1+\eta_{i}%
)\delta(u)\right]  \right\}
\end{align}
where $p^{2}=P^{2}/\Lambda^{2}$. Note that if $w=n$ then $w\cdot n=0$ and we
get $\mathcal{J}_{0}$ of eq. (\ref{J0end}) multiplied by $uP^{+}$ as it should
be, since we could have used $\delta\left(  k\cdot n-u\,n\cdot P\right)  $ in
the first place. Similarly if $w=\varepsilon_{\bot}$ we have $\mathcal{J}%
_{1}=0$ which means that a single power of $\kappa_{\bot}$ integrates to zero.
Note that due to Lorentz invariance after $du$ integration the coefficient in
front of $w\cdot n$ should vanish in accordance with (\ref{LorentzB0}).

Finally if the numerator contains $\kappa^{\mu}\kappa^{\nu}$, let's call such
an integral $\mathcal{J}_{2}$, we have%
\begin{equation}
\mathcal{N\rightarrow N}(\eta_{i},\eta_{j})(w\cdot\kappa)(v\cdot
\kappa)=\mathcal{N}(\eta_{i},\eta_{j})(\vec{w}\cdot\vec{\kappa})(\vec{v}%
\cdot\vec{\kappa}).
\end{equation}
Using (\ref{indel}) and integrating over $dy$ we get three different
contributions to $\mathcal{J}_{2}$ depending on the tensor structure:%
\begin{align}
\mathcal{J}_{2}^{(0)} &  =i\frac{\mathcal{A}}{16\pi^{2}}\frac{\Lambda^{2}%
}{P^{+}}\left(  \frac{e^{-\gamma}}{4\pi\Lambda^{2}}\right)  ^{-\varepsilon
}\frac{1}{\varepsilon}%
{\displaystyle\sum\limits_{i,j=1}^{4n+1}}
f_{i}\,f_{j}\,\frac{\eta_{i}^{4n}\eta_{j}^{4n}\,\mathcal{N}}{\left[  1-\bar
{u}\eta_{i}+u\eta_{j}+u\bar{u}\,p^{2}\right]  ^{\varepsilon}}\nonumber\\
&  \qquad\left\{  -\frac{1}{2}\frac{\left[  1-\bar{u}\eta_{i}+u\eta_{j}%
+u\bar{u}\,p^{2}\right]  }{(\varepsilon-1)}\left(  w\cdot v\right)
+\frac{u^{2}}{\Lambda^{2}}\left(  v\cdot P\right)  \left(  w\cdot P\right)
\right\}  ,\label{J20end}%
\end{align}%
\begin{align}
\mathcal{J}_{2}^{(1)} &  =i\frac{\mathcal{A}}{16\pi^{2}}\frac{\Lambda^{2}%
}{2P^{+\,2}}\left(  \frac{e^{-\gamma}}{4\pi\Lambda^{2}}\right)  ^{-\varepsilon
}\frac{1}{\varepsilon}\left\{  \left(  w\cdot P\right)  \left(  v\cdot
n\right)  \frac{{}}{{}}+\left(  w\cdot n\right)  \left(  v\cdot P\right)
\right\}  \nonumber\\
&
{\displaystyle\sum\limits_{i,j=1}^{4n+1}}
f_{i}\,f_{j}\,\frac{\eta_{i}^{4n}\eta_{j}^{4n}\,\mathcal{N}}{\left[  1-\bar
{u}\eta_{i}+u\eta_{j}+u\bar{u}\,p^{2}\right]  ^{\varepsilon}}\left\{  \left(
1+\eta_{j}\right)  \delta(u-1)\frac{{}}{{}}\right.  \nonumber\\
&  \,\qquad\left.  -\left(  1-\bar{u}\eta_{i}+u\eta_{j}+u\bar{u}%
\,p^{2}\right)  +\,u\left[  \left(  \eta_{i}-\eta_{j}\right)  +\frac{{}}{{}%
}(1-2u)p^{2}\right]  \right\}  \label{J21end}%
\end{align}
and finally%
\begin{align}
\mathcal{J}_{2}^{(2)} &  =i\frac{\mathcal{A}}{16\pi^{2}}\frac{\Lambda^{4}%
}{4P^{+\,3}}\left(  \frac{e^{-\gamma}}{4\pi\Lambda^{2}}\right)  ^{-\varepsilon
}\frac{1}{\varepsilon}\left(  w\cdot n\right)  \left(  v\cdot n\right)
{\displaystyle\sum\limits_{i,j=1}^{4n+1}}
f_{i}\,f_{j}\,\frac{\eta_{i}^{4n}\eta_{j}^{4n}\mathcal{N}}{\left[  1-\bar
{u}\eta_{i}+u\eta_{j}+u\bar{u}\,p^{2}\right]  ^{\varepsilon}}\nonumber\\
&  \left\{  \frac{1}{2}\left[  \left[  1+\eta_{j}\right]  ^{2}\delta^{\prime
}(u-1)\frac{{}}{{}}-\left[  1+\eta_{i}\right]  ^{2}\delta^{\prime}(u)\right]
\right.  \nonumber\\
&  \qquad+\left[  \left(  \eta_{i}-\eta_{j}\right)  +\frac{{}}{{}}%
(1-2u)p^{2}\right]  \left[  \left[  1+\eta_{j}\right]  \delta(u-1)\frac{{}}%
{{}}-\left[  1+\eta_{i}\right]  \delta(u)\right]  \nonumber\\
&  \qquad\qquad\qquad\left.  +\left[  1-\bar{u}\eta_{i}+u\eta_{j}+u\bar
{u}\,p^{2}\right]  \,up^{2}+\left[  \left(  \eta_{i}-\eta_{j}\right)
+\frac{{}}{{}}(1-2u)p^{2}\right]  ^{2}\right\}  .\nonumber\\
&  \;\label{J22}%
\end{align}
In eq. (\ref{J22}) we encounter derivatives of $\delta$ functions; it is here
implicitly assumed that the coefficient $\mathcal{N}\left[  1-\bar{u}\eta
_{i}+u\eta_{j}+u\bar{u}\,p^{2}\right]  ^{-\varepsilon}$ when multiplied by
$\delta^{\prime}(u-1)$ or $\delta^{\prime}(u)$ is taken at the corresponding
value of $u$. Note that Lorentz invariance requires that
\begin{equation}%
{\displaystyle\int\limits_{0}^{1}}
du\,\mathcal{J}_{2}^{(1,2)}=0
\end{equation}
(modulo possible subtraction of the perturbative part).

Finally let us remark that if we need an integral of $k_{\bot}^{2}$ we may use
the following trick in two dimensional transverse plane:%
\begin{equation}%
{\displaystyle\int}
d^{2}\vec{k}_{\bot}\,\vec{k}_{\bot}^{\,2}=2%
{\displaystyle\int}
d^{2}\vec{k}_{\bot}\,(\vec{\varepsilon}_{\bot}\cdot\vec{k}_{\bot})^{2}
\label{ktint}%
\end{equation}
if there is no other dependence on the transverse angle, as it indeed happens
in our case. We can then evaluate the r.h.s of eq. (\ref{ktint}) using the
formulae from the present appendix.

\section{Axial form factor}

\label{axialf}

In this appendix we show, as an example, simple calculation of the axial form
factor. We start from the matrix element on the left hand side in eq.
(\ref{eq:ax_ff}):%
\begin{align}
\left\langle 0\left\vert \overline{\psi}\left(  -\lambda n\right)  \gamma
^{\mu}\gamma_{5}\psi\left(  \lambda n\right)  \right\vert \gamma\left(
P,\varepsilon\right)  \right\rangle  &  =-eN_{c}\varepsilon_{\nu}\int
\frac{d^{D}k}{\left(  2\pi\right)  ^{D}}\,e^{i\left(  2k\cdot n-P\cdot
n\right)  \lambda}\\
&  \operatorname*{Tr}\left\{  \gamma^{\mu}\gamma_{5}\frac{1}{\not k  -M_{k}%
}\gamma^{\nu}\frac{1}{(\not k  -\not P  )-M_{k-P}}\right\}  .\nonumber
\end{align}
Calculating the trace and taking the derivative with respect to $\lambda$ as
in the definition (\ref{eq:ax_ff}) we obtain%
\[
\mathcal{M}\equiv\frac{d}{d\lambda}\left.  \left\langle 0\left\vert
\overline{\psi}\left(  -\lambda n\right)  \gamma^{\mu}\gamma_{5}\psi\left(
\lambda n\right)  \right\vert \gamma\left(  P,\varepsilon\right)
\right\rangle \right\vert _{\lambda=0}=4eN_{c}\varepsilon_{\nu}P_{\beta
}\varepsilon^{\mu\nu\alpha\beta}\int\frac{d^{D}k}{\left(  2\pi\right)  ^{D}%
}\,\frac{k_{\alpha}\left(  2k\cdot n-P\cdot n\right)  }{D\left(  k\right)
D\left(  k-P\right)  }.
\]
Using Lorentz invariance and some simple algebra we get%
\begin{equation}
\mathcal{M}=-2eN_{c}\varepsilon_{\nu}P_{\alpha}n_{\beta}\varepsilon^{\mu
\nu\alpha\beta}\int\frac{d^{D}k}{\left(  2\pi\right)  ^{D}}\,\frac{\left(
2k\cdot n-P\cdot n\right)  \left(  k\cdot\tilde{n}-\bar{p}k\cdot n\right)
}{D\left(  k\right)  D\left(  k-P\right)  } \label{eq:aff3}%
\end{equation}
with $\bar{p}=P^{2}/P^{+\,2}$. Comparing this with the right hand side of
(\ref{eq:ax_ff}) we get the following expression%
\begin{equation}
F_{A}\left(  P^{2}\right)  =\frac{2iN_{c}}{f_{3\gamma}}\,\mathcal{J},
\label{eq:aff4}%
\end{equation}
where $\mathcal{J}$ denotes the integral in \eqref{eq:aff3}. However it can be
easily shown that Lorentz invariance requires that
\begin{equation}
\int\frac{d^{D}k}{\left(  2\pi\right)  ^{D}}\,\frac{k\cdot n\,k\cdot\tilde
{n}-\bar{p}\,\left(  k\cdot n\right)  ^{2}}{D\left(  k\right)  D\left(
k-P\right)  }=\frac{2}{2-D}\int\frac{d^{D}k}{\left(  2\pi\right)  ^{D}}%
\,\frac{k_{T}^{2}}{D\left(  k\right)  D\left(  k-P\right)  } \label{eq:aff5}%
\end{equation}
and%
\begin{equation}
\int\frac{d^{D}k}{\left(  2\pi\right)  ^{D}}\,\frac{k\cdot\tilde{n}-\bar
{p}\,k\cdot n}{D\left(  k\right)  D\left(  k-P\right)  }=0. \label{eq:aff6}%
\end{equation}
Using this, $\mathcal{J}$ reduces to%
\begin{equation}
\mathcal{J}=\frac{4}{2-D}\int\frac{d^{D}k}{\left(  2\pi\right)  ^{D}}%
\,\frac{k_{T}^{2}}{D\left(  k\right)  D\left(  k-P\right)  }. \label{eq:aff7}%
\end{equation}
This integral reads%
\begin{equation}
\mathcal{J}=\frac{\Lambda^{D-2}}{p^{+}}\,\frac{4i}{D-2}\left(  \frac{1}{4\pi
}\right)  ^{D/2}\,\sum_{i,j=1}^{4n+1}f_{i}f_{j}\eta_{i}^{4n}\eta_{j}^{4n}%
\int_{0}^{1}du\left(  1-\bar{u}\eta_{i}+u\eta_{j}+u\bar{u}\,p^{2}\right)
^{1-\epsilon}\Gamma\left(  \epsilon\right)  . \label{eq:aff8}%
\end{equation}
Expanding in $\epsilon$ and subtracting the perturbative part we obtain the
final expression%
\begin{align}
F_{A}\left(  P^{2}\right)   &  =\frac{1}{4\pi^{2}}\,\frac{\Lambda^{2}N_{c}%
}{f_{3\gamma}}\,\sum_{i,j=1}^{4n+1}f_{i}f_{j}\eta_{i}^{4n}\eta_{j}%
^{4n}\label{eq:aff9}\\
&  \int_{0}^{1}du\,\left(  1-\bar{u}\eta_{i}+u\eta_{j}+u\bar{u}\,p^{2}\right)
\ln\left(  \frac{1-\bar{u}\eta_{i}+u\eta_{j}+u\bar{u}\,p^{2}}{u\bar{u}p^{2}%
}\right)  .\nonumber
\end{align}
Integration over $du$ can, in principle, be done analytically (taking into
account remarks given in the main text), however here we just plot
$F_{A}\left(  P^{2}\right)  $ in sect. \ref{axialh}.

\end{document}